\documentclass[fleqn,usenatbib]{mnras}

\usepackage{newtxtext,newtxmath}
\usepackage{caption}
\usepackage{threeparttable}
\usepackage[T1]{fontenc}

\DeclareRobustCommand{\VAN}[3]{#2}
\let\VANthebibliography\thebibliography
\def\thebibliography{\DeclareRobustCommand{\VAN}[3]{##3}\VANthebibliography}

\usepackage{graphicx}	% Including figure files
\usepackage{amsmath}	% Advanced maths commands

\title[Compositions of the Hercules-Aquila Cloud and Virgo Over-density]{Compositions of the Hercules-Aquila Cloud and Virgo Over-density}

\author[]{
Dashuang Ye,$^{1}$
Cuihua Du,$^{1}$\thanks{E-mail: ducuihua@ucas.ac.cn}
Mingji Deng,$^{1}$
Jiwei Liao,$^{1}$
Yang Huang,$^{1,2}$
Jianrong Shi,$^{2,1}$
Jun Ma,$^{2,1}$
\\
$^{1}$College of Astronomy and Space Sciences, University of Chinese Academy of Sciences, Beijing 100049, P.R. China\\
$^{2}$Key Laboratory of Optical Astronomy, National Astronomical Observatories, Chinese Academy of Sciences, Beijing 100012, P.R.China
}

\date{Accepted XXX. Received YYY; in original form ZZZ}

\pubyear{2015}

\begin{document}
\label{firstpage}
\pagerange{\pageref{firstpage}--\pageref{lastpage}}
\maketitle

\begin{abstract}
Based on a sample of K giant from Large sky Area Multi-Object fiber Spectroscopic Telescope (LAMOST) Data Release 8 and a sample of RR Lyrae (RRL) from \textit{Gaia} Data Release 3, we investigate the compositions of the Hercules-Aquila Cloud (HAC) and Virgo Over-density (VOD) and their collective contribution to the tilt and triaxiality of the stellar halo ($r\,\textless\,40\,{\rm kpc}$) as well as two breaks at $\approx15\,{\rm kpc}$ and 30\,kpc.
We apply the Gaussian mixture model (GMM) to divide the stellar halo into the isotropic component and the radially biased anisotropic component, namely Gaia-Sausage-Enceladus (GSE), and find that both HAC and VOD are dominated by the GSE debris stars with weights of $0.67^{+0.09}_{-0.07}$ and $0.57^{+0.07}_{-0.06}$, respectively.
In addition, using the K giants with orbital parameters, we identify the member stars of known substructures, including GSE, Sagittarius (Sgr), Helmi Streams, Sequoia, Thamnos, Pontus, Wukong, and Metal-weak Thick Disk (MWTD), to probe the compositions of low-eccentricity stars in the HAC and VOD regions.
In density fittings of the RRL sample, we note that the absence of HAC and VOD has a weak effect on the shape of halo.
Finally, we find that the radially biased anisotropic halo contributes majorly to the stellar halo that can be modelled with a tilted triaxial ellipsoid and a doubly broken power law with breaking radii at $18.08^{+2.04}_{-3.22}\,{\rm kpc}$ and $33.03^{+1.30}_{-1.21}\,{\rm kpc}$.
This has important significance for understanding the status of large diffuse over-densities in the Milky Way.
\end{abstract}


%%%%%%%%%%%%%%%%%%%%%%%%%%%%%%%%%%%%%%%%%%%%%%%%%%

%%%%%%%%%%%%%%%%% BODY OF PAPER %%%%%%%%%%%%%%%%%%

\section{Introduction}

According to the $\rm \Lambda CDM$-based models of galaxy formation, the stellar halo of the Milky Way (MW) was primarily formed by a series of accretion and merger events \citep[][]{Searle1978ApJ,Blumenthal1984Natur,Bullock2005ApJ}.
Observationally, several mergers have been discovered in the Galactic stellar halo thanks to the \textit{Gaia} mission \citep{Gaia2016A&A} and stellar spectroscopic surveys such as Apache Point Observatory Galactic Evolution Experiment \citep[APOGEE;][]{apogee2017AJ,Abdurro2022ApJS}, Galactic Archaeology with High Efficiency and Resolution Multi-Element Spectrograph \citep[GALAH;][]{GALAH2015MNRAS,Martell2017MNRAS,Buder2021MNRAS}, LAMOST \citep{Zhao2006ChJAA,zhao2012lamost,cui2012large}, and Sloan Exploration of Galactic Understanding and Evolution \citep[SEGUE;][]{segue2009AJ,Alam2015ApJS}.
One of the significant accretion events is a single dwarf galaxy named as Gaia-Sausage-Enceladus that merged with the MW 8-10 Gyr ago \citep[e.g.,][]{Haywood2018ApJ,Gallart2019NatAs} and contributed the bulk of the inner halo \citep[e.g.,][]{Belokurov2018MNRAS,Helmi2018Natur,Naidu2020ApJ,iorio2021MNRAS,lancaster2019MNRAS,Wu2022ApJ}.
Also, some unremarkable substructures, such as slightly retrograde Thamnos \citep{Koppelman2019A&A}, Sequoia \citep{Myeong2019MNRAS} and the metal-poor Wukong/LMS-1 \citep{Naidu2020ApJ,Yuan2020ApJ}, were found to be separated in the integrals of motion space, but it is currently unclear whether they originated from a progenitor galaxy different from that of GSE.

Two large and spatially diffuse cloud-like structures, namely the Hercules-Aquila Cloud \citep[HAC,][]{Besla2007ApJ,Belokurov2007ApJ} and the Virgo Over-density \citep[VOD,][]{Vivas2001ApJ,Newberg2002ApJ}, have been interpreted as apocentric pile-ups \citep{Deason2018ApJ,naidu2021ApJ,Ye2023MNRAS}.
\citet{naidu2021ApJ} ran a grid of $\sim500$ \textit{N}-body galaxy mergers and identified a plausible configuration of GSE that reproduces several key properties of the merger remnant, including HAC and VOD, two breaks in the radial profile, and a large-scale tilt in the stellar halo.
\citet{Han2022ApJ} found that HAC and VOD are long-lived structures associated with GSE under a tilted halo potential.
In conclusion, it has been shown that GSE is inextricably linked to HAC and VOD.

Observationally, recent studies suggest that the HAC and VOD, with a very similar range of distances, have a common origin \citep{Simion2019MNRAS} and have been linked to the GSE \citep{Perottoni2022ApJ}.
The HAC extends in heliocentric distances ($d_{\odot}$) from 10 to 20\,kpc, with $25^{\circ}\,\textless\,l\,\textless\,60^{\circ}$ and $-40^{\circ}\,\textless\,b\,\textless\,40^{\circ}$ \citep{Belokurov2007ApJ}.
Although the distribution of VOD is clearly distinct from that of HAC in spatial locations of $270^{\circ}\,\textless\,l\,\textless\,330^{\circ}$ and $50^{\circ}\,\textless\,b\,\textless\,75^{\circ}$ \citep{Juri2008ApJ,Bonaca2012AJ}, they have similar orbital properties and chemical abundances, and are well mixed in velocity space.
\citet{Simion2019MNRAS} and \citet{Yan2023A&A} have shown the density maps of the member stars of HAC and VOD along their individual orbits over the past 8 Gyr, and found that they show extremely similar distributions.
\citet{Perottoni2022ApJ} combined SEGUE, APOGEE, \textit{Gaia}, together with \textbf{\tt\string StarHorse} distances to investigate the potential link between the HAC and VOD as well as the GSE, and found that, similar to the GSE, the VOD and HAC are composed mostly of stars with $e\,\textgreater\,0.7$, and that the HAC and VOD are indistinguishable from the prototypical GSE population within all chemical-abundance spaces (e.g., [Mg/Fe]-[Fe/H], [Mg/Mn]-[Al/Fe], [Al/Fe]-[Fe/H], and [Ni/Fe]-[(C+N)/O] planes).
\citet{Perottoni2022ApJ} also identified some metal-poor stars ($\rm [Fe/H]\,\textless\,-1.5$) with mildly eccentric orbits ($e\sim0.5$), they might originate from the metal-poor portion of GSE or other accretion events.

In this work, we use K giants and RR Lyrae as tracers to investigate the GSE weight of HAC (VOD) and the existence of other substructures as well as the difference in density profile of the stellar halo before and after removing the RRab stars from HAC and VOD.
In Section \ref{Data}, we present the data sets and selection criteria used to select the parent sample.
In Section \ref{Methods}, we describe the Gaussian mixture model for the RRL sample, the selection criteria adopted to identify known substructures, as well as density fitting.
In Section \ref{Results}, we analyze the GMM fitting results for the RRL sample, the orbital properties of all substructures for K giants, and density fitting results for three RRL datasets.
Finally, a brief summary is given in Section \ref{Conclusions}.

\section{Data}\label{Data}
\subsection{K giant sample}\label{K giant sample}

\begin{figure}
	\includegraphics[width=\columnwidth]{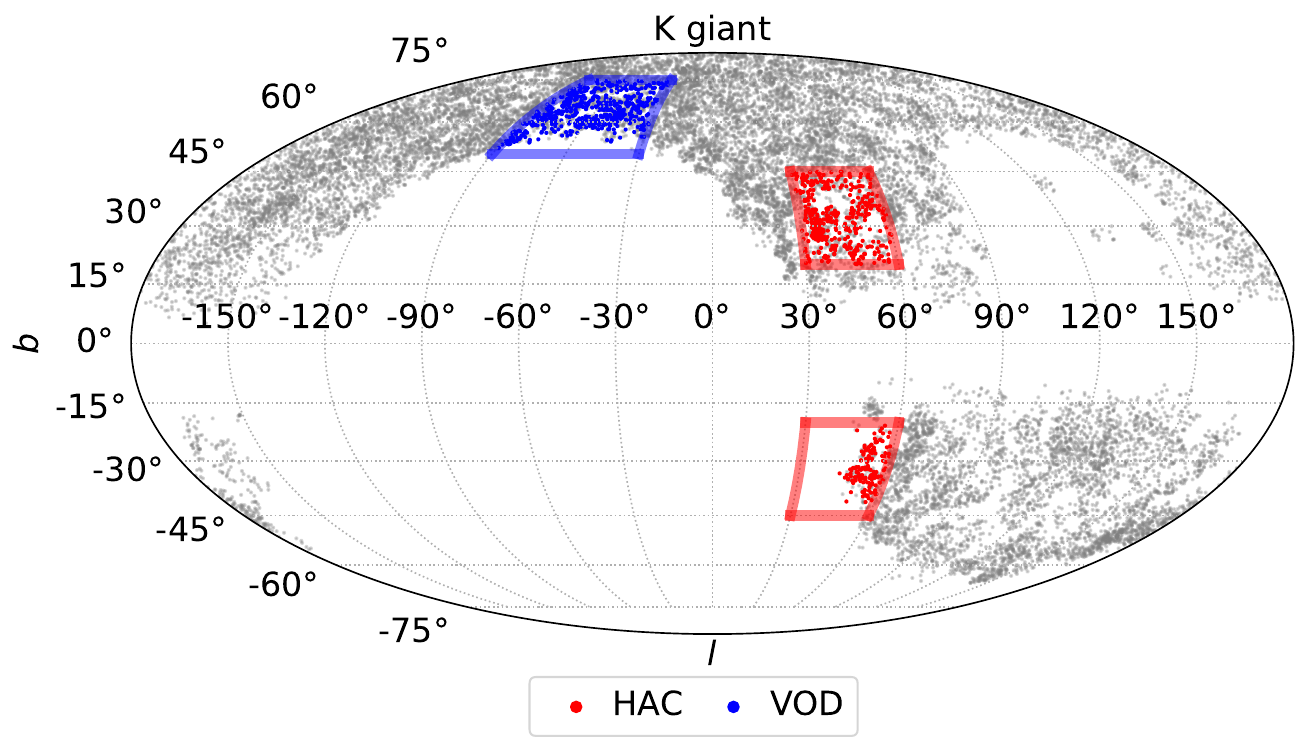}
    \caption{Spatial projection in Galactic coordinates of K giants from HAC (red), VOD (blue), and total K giant sample (grey).}
    \label{figwholeK}
\end{figure}

\begin{figure}
	\includegraphics[width=\columnwidth]{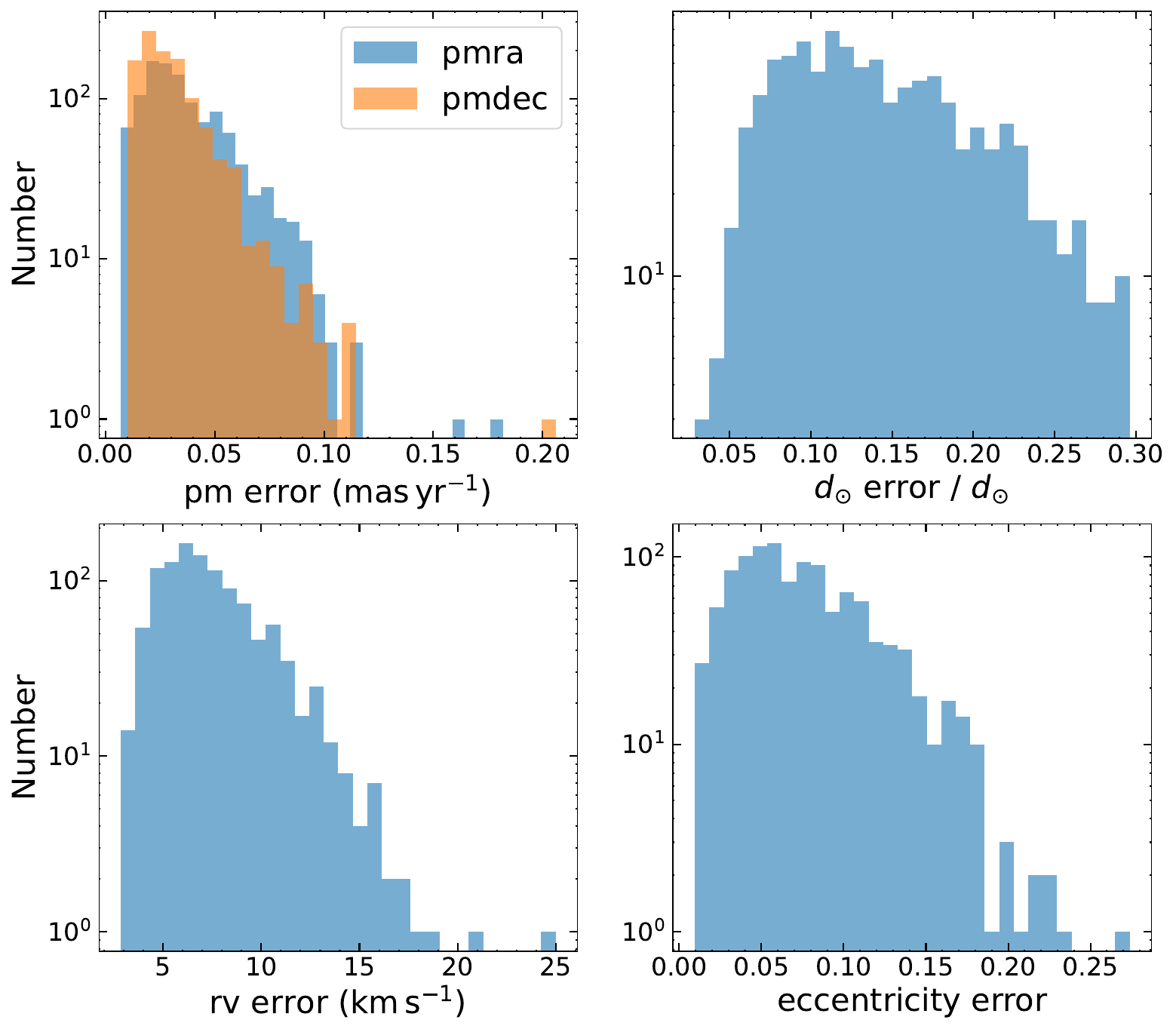}
    \caption{Errors in the HAC and VOD data sets selected in the K giant sample here. Top-left panel: the distribution of proper motion errors. Note that standard errors of proper motions in right ascension (steel blue) and declination directions (orange) from \textit{Gaia} DR3 are almost less than 0.1 mas $\rm yr^{-1}$. Top-right panel: the distribution of distance relative errors from \citet{zhanglan2023}. Bottom-left panel: the distribution of radial velocity errors from LAMOST DR8. Bottom-right panel: the distribution of eccentricity errors.}
    \label{figdataerror}
\end{figure}

\begin{figure}
	\includegraphics[width=0.9\columnwidth]{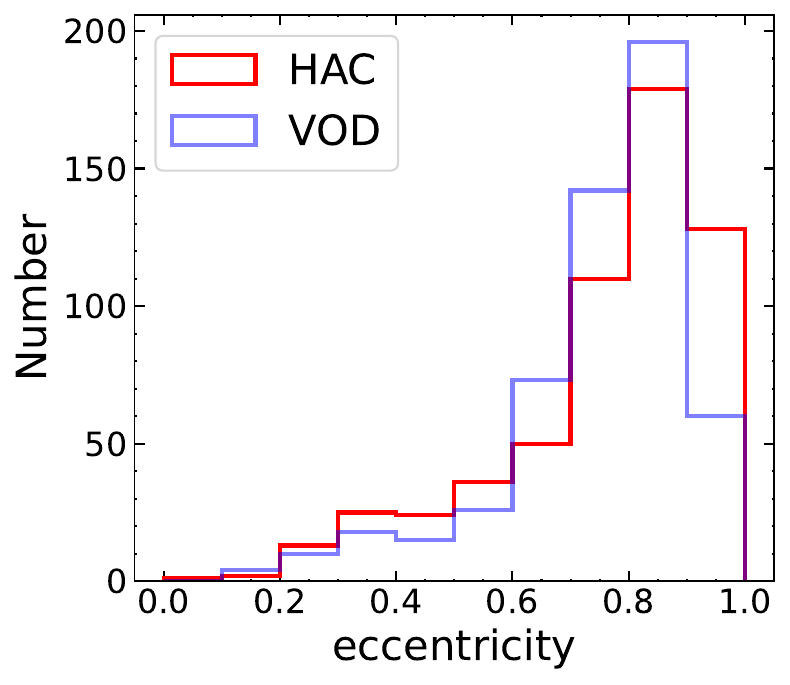}
    \caption{Orbital eccentricity distribution of K giants from HAC (red) and VOD (steel blue).}
    \label{fighacvod_e}
\end{figure}

\citet{zhanglan2023} recently provided a publicly available catalog of distances for $19,544$ K giants drawn from LAMOST DR8 \citep{Zhao2006ChJAA,zhao2012lamost,cui2012large}.
They estimated the absolute magnitudes in the SDSS \textit{r} band, the distance moduli, and the corresponding uncertainties through a Bayesian approach devised by \citet{Xue2014ApJ}.
The stars in this catalog lie in a region of $4-126\,$kpc from the Galactic center, forming the largest spectroscopic sample of distant tracers in the Milky Way halo so far.
By cross-matching ($1^{\prime\prime}$ search radius) with \textit{Gaia} Data Release 3 \citep[\textit{Gaia} DR3,][]{GaiaDR32023A&A}, we obtained the proper motions and their uncertainties.
We correct the line-of-sight velocities ($v_{\rm los}$) of LAMOST DR8 by adding the radial velocity offset ($\mu=6.1\,{\rm km\,s^{-1}}$) between LAMOST DR8 and \textit{Gaia}, in which we remove stars with $\lvert v_{\rm los,\,\textit{Gaia}}-v_{\rm los,\,LAMOST}-\mu\rvert\,\textgreater\,3\,\sigma$ ($\sigma=9.6\,{\rm km\,s^{-1}}$) to avoid mismatches.
We apply the selection criteria of \citet{Perottoni2022ApJ} to extract the stars from HAC and VOD, in which the HAC is divided into the HAC-South (HAC-S) and HAC-North (HAC-N).
The selection criteria of HAC-S and HAC-N are as follows: $30^{\circ}\,\textless\,l\,\textless\,60^{\circ}$ and $-45^{\circ}\,\textless\,b\,\textless-20^{\circ}$, $30^{\circ}\,\textless\,l\,\textless\,60^{\circ}$ and $20^{\circ}\,\textless\,b\,\textless\,45^{\circ}$.
For the VOD, our selection is based on the following Galactic coordinate cuts: $270^{\circ}\,\textless\,l\,\textless\,330^{\circ}$ and $50^{\circ}\,\textless\,b\,\textless75^{\circ}$.
For both overdensities, we consider heliocentric distances ($d_{\odot}$) between 10 and 20\,kpc. 
The above selection criteria yield 568 and 544 K giants from HAC and VOD, respectively.

In Figure \ref{figwholeK}, we show the HAC (red) and VOD (blue) stars as well as the total sample of K giant stars (grey), in which the relative errors of distance are almost ($\textgreater\,99\%$) less than 0.3, and they also have accurate radial velocities from LAMOST. 
Combined with the precise proper motions of \textit{Gaia} DR3, we construct a full 7D data set consisting of positions, space motions, and metallicity, and this sample is very suitable for studying the local kinematics of stellar halo.
However, as seen in Figure \ref{figwholeK}, the observations of LAMOST do not completely cover the fields of HAC and VOD.
In Figure \ref{figdataerror}, we show the distributions of errors of HAC and VOD member stars in the total K giant sample, including the errors of proper motions, the relative errors of distance, the errors of line-of-sight velocity, and the eccentricity errors.
Note that the errors of proper motions in both directions are almost less than 0.1\,mas\,$\rm yr^{-1}$, which corresponds to an error of $9.5\,\rm km\,s^{-1}$ at $d_{\odot}=20\,{\rm kpc}$.
Considering the median error of $7\,{\rm km\,s^{-1}}$ for line-of-sight velocity and relative distance errors less than 0.3, the HAC and VOD data sets allow us to accurately measure their kinematics and orbits.

We adopt a right-handed Galactocentric frame of reference: \textit{x} and \textit{z} are positive in the directions of the Sun and the North Galactic Pole (NGP), and the $y$-axis is opposite to the direction of the Galactic rotation.
The Sun is located at $(R_{\odot},\,0,\,z_{\odot})=(8.2,\, 0,\, 0.025)\,$kpc \citep{Juri2008ApJ,bland2016ARA&A}.
We calculate the Galactic space velocities from heliocentric distances, proper motions, and line-of-sight velocities \citep{johnson1987AJ}.
We correct the velocities based on the solar peculiar motion $(U_{\odot},\,V_{\odot},\,W_{\odot})=(10.,\,11.,\,7.)\,{\rm km\,s^{-1}}$ \citep{tian2015ApJ,bland2016ARA&A} and the local standard of rest (LSR) velocity $V_{\rm LSR}=232.8\,{\rm km\,s^{-1}}$ \citep{McMillan2017MNRAS}. $U,\,V$ and $W$ point toward the Galactic center, the Galactic rotation, and the NGP, respectively.
$v_r,\,v_{\theta},\,v_{\phi}$ represent the spherical radial, zenithal, and azimuthal velocities, respectively.
Note that $R$ and $v_R$ are the cylindrical radius and the cylindrical radial velocity. In order to transform to the Galactocentric spherical and cylindrical coordinate systems, we follow Equations (\ref{tranformxyz}) and (\ref{tranformvxyz}):
\begin{equation}\label{tranformxyz}
\begin{aligned}
r &= \sqrt{x^2+y^2+z^2}\\
\theta &= \pi/2-\text{arctan}(z/\sqrt{x^2+y^2})\\
\phi &= \text{arctan2}(y/x)\\
R &= \sqrt{x^2+y^2},
\end{aligned}
\end{equation}
where $\text{tan}\,(\text{arctan2}\,(y,\,x))=y/x$,

\begin{equation}\label{tranformvxyz}
\begin{aligned}
\begin{bmatrix}
    v_r\\
    v_{\theta}\\
    v_{\phi}\\
\end{bmatrix} &= \textbf{U}_{\rm s}
\begin{bmatrix}
    v_x\\
    v_y\\
    v_z\\
\end{bmatrix}\\
\begin{bmatrix}
    v_R\\
    v_{\phi}\\
    v_{z}\\
\end{bmatrix} &= \textbf{U}_{\rm c}
\begin{bmatrix}
    v_x\\
    v_y\\
    v_z\\
\end{bmatrix},
\end{aligned}
\end{equation}
where
\begin{equation}
\begin{aligned}
\textbf{U}_{\rm s} &= 
\begin{bmatrix}
    \text{sin}\,\theta\,\text{cos}\,\phi & \text{sin}\,\theta\,\text{sin}\,\phi & \text{cos}\,\theta\\
    \text{cos}\,\theta\,\text{cos}\,\phi & \text{cos}\,\theta\,\text{sin}\,\phi & -\text{sin}\,\theta \\
    -\text{sin}\,\phi & \text{cos}\,\phi & 0
\end{bmatrix}\\
\textbf{U}_{\rm c} &=
\begin{bmatrix}
    \text{cos}\,\phi & \text{sin}\,\phi & 0\\
    -\text{sin}\,\phi & \text{cos}\,\phi & 0\\
    0 & 0 & 1
\end{bmatrix}.
\end{aligned}
\end{equation}
To compute the orbits of stars from the HAC and VOD, we adopt the axisymmetric Galactic potential model of \citet{McMillan2017MNRAS}. 
We calculate the actions using the Python package \textbf{\tt\string galpy} \citep{Bovy2015ApJSgalpy} and St$\ddot{\text{a}}$ckel approximation of \citet{Binney2012MNRAS}, and integrate the orbits forward in a time scale of 10 Gyr to obtain vital orbital parameters including eccentricity ($e$), apocentric and pericentric distances ($r_{\rm apo}$ and $r_{\rm peri}$), the maximum height from the Galactic plane ($z_{\rm max}$).
For each star, we construct a set of 100 initial conditions using a Monte Carlo technique to propagate the errors in heliocentric distances, proper motions, and line-of-sight velocities.
In the bottom-right panel of Figure \ref{figdataerror}, the eccentricity errors are generally less than 0.15, with a median value of only 0.07, so the distributions of eccentricities of K giants from HAC (red) and VOD (steel blue) are sufficiently reliable.
As shown in Figure \ref{fighacvod_e}, both eccentricity distributions have a peak around $e\sim0.8$ with a width of 0.1 and a low-eccentricity tail ($e\,\textless\,0.6$), which resemble those in \citet{Perottoni2022ApJ}.

\subsection{RR Lyrae sample}

\begin{figure}
	\includegraphics[width=\columnwidth]{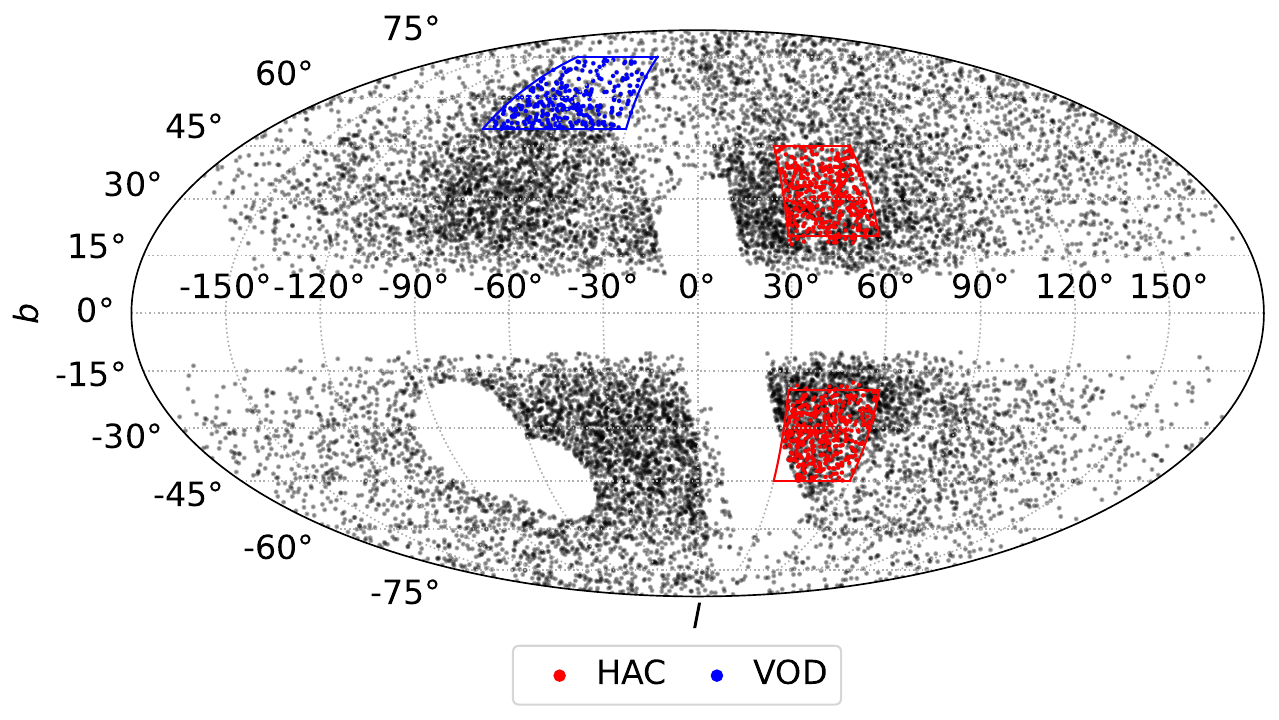}
    \centering
    \caption{Spatial projection in Galactic coordinates of RRLs from HAC (red), VOD (blue), and our RRL sample (black).}
    \label{figRRLall}
\end{figure}

On the basis of pulsating modes, RR Lyrae stars are divided into three types, namely RRab, RRc, and RRd.
\citet{GaiaDR32023A&A} have published 270,\,905 RRLs (174,\,947 RRab stars, 93,\,952 RRc stars, and 2,\,006 RRd stars), processed by the dedicated Specific Objects Study (SOS) Cep\&RRL pipeline \citep{Clementini2023A&A}. 
Here, the RRab stars from \textit{Gaia} DR3 are used as our initial sample due to their high completeness. 
However, the radial velocity measurements for most of the stars are lacking due to the periodic radial expansion/contraction of the RRL surface layers.
As mentioned in Section \ref{K giant sample}, the K giant sample has reliable 7D information but its spatial completeness is not enough, so both samples complement each other.

RR Lyrae stars are ideal standard candles, thanks to their well-defined absolute magnitude-metallicity relationship in the visible band and the period-absolute magnitude-metallicity relationship in the near/mid-infrared.
On the basis of 2,\,046 RRab stars, with accurate pulsation periods (\textit{P}), Fourier parameters ($\phi_{31}$, $R_{21}$) and metallicity measurements ([Fe/H]) from \textit{Gaia} mission, LAMOST, and the Sloan Digital Sky Surve, \citet{Lixinyi2023ApJ} gave the new $P\text{--}\phi_{31}\text{--}R_{21}\text{--}{\rm [Fe/H]}$ and $M_{\rm G}\text{--}{\rm [Fe/H]}$  relationships for RRab stars, as follows:
\begin{equation}
\begin{aligned}
\text{[Fe/H]} &= (-1.888\pm0.002) + (-5.772\pm0.026)(P-0.6) \\
&+ (1.090\pm0.005)(\phi_{31}-2) + (1.065\pm0.030)(R_{21}-0.45)\\
M_{\rm G} &= (0.350\pm0.016)\text{[Fe/H]}+(1.106\pm0.021).
\end{aligned}
\end{equation}
In case the star has not a period or $\phi_{31}$ and $R_{21}$, these values are drawn from their overall 3D distribution considering the whole \textit{Gaia} SOS catalogue making use of \text{\tt\string XDGMM} \citep{Bovy2011AnApS} in the Python package \text{\tt\string astroML} \citep{astroML,astroMLText} following \citet{iorio2021MNRAS}.
We correct the observed \textit{G}-band magnitude ($ G_{\rm obs}$) as:
\begin{equation}
\begin{aligned}
G &= G_{\rm obs}-k_{\rm G}E(B-V),
\end{aligned}
\end{equation}
where $E(B-V)$ and its error, $\delta_{E(B-V)}=0.16\times E(B-V)$, come from SFD model \citep{SFD1998ApJ}, assuming $k_{\rm G}=2.516$ \citep{Huang2021ApJ}. Furthermore, we produce 1,\,000 realizations of the heliocentric distance using the equation
\begin{equation}
\begin{aligned}
\text{log}\,\frac{d_{\odot}}{{\rm kpc}} = \frac{G-M_{\rm G}}{5}-2,
\end{aligned}
\end{equation}
and they are used to propagate errors to 6D information, including 3D spatial positions, 2D tangential velocities, and metallicity.
As described in Section \ref{K giant sample}, we gain 3D spatial information in Cartesian coordinates, spherical coordinates, and cylindrical coordinates.
We estimate the 2D tangential velocities from the observed proper motions as follows:
\begin{equation}
\begin{aligned}
v_{l} &= 4.74057\,\mu_{l}\,d_{\odot} + v_{l,\odot}\\
v_{b} &= 4.74057\,\mu_{b}\,d_{\odot} + v_{b,\odot}.\\
\end{aligned}
\end{equation}
Note that we use the right-hand system here, and we estimate $v_{l,\odot}$ and $v_{b,\odot}$, namely the projection of the Sun velocity $\textbf{v}_{\odot}=(v_{x,\odot},\,v_{y,\odot},\,v_{z,\odot})=(-10.,\,-243.8,\,7.)\,{\rm km\,s^{-1}}$ in the tangential plane at the position of the star, by transformation:
\begin{equation}
\begin{aligned}
\begin{bmatrix}
v_{b}\\
v_{l}\\
v_{\rm los}\\
\end{bmatrix}
= \textbf{U}_1
\begin{bmatrix}
v_{\rm x}\\
v_{\rm y}\\
v_{\rm z}\\
\end{bmatrix},
\end{aligned}
\end{equation}
where
\begin{equation}
\begin{aligned}
\textbf{U}_1=
\begin{bmatrix}
\text{sin}\,b\,\text{cos}\,l & \text{sin}\,b\,\text{sin}\,l & \text{cos}\,b\\
\text{sin}\,l & -\,\text{cos}\,l & 0\\
-\,\text{cos}\,b\,\text{cos}\,l & -\,\text{cos}\,b\,\text{sin}\,l & \text{sin}\,b\\
\end{bmatrix}.
\end{aligned}
\end{equation}

Following \citet{iorio2021MNRAS}, given the significant increase in velocity uncertainties at large distances, we also limit the extent of our sample to within 40\,kpc from the Galactic centre. Compact overdensities have a significant impact on break radius \citep{Xue2015ApJ}, and the extinction correction in the areas of Magellanic Clouds is not good enough, which will lead to inaccurate distances. 
In particular, overdensities will also affect kinematic fit \citep{iorio2021MNRAS}.
Therefore, we need to spatially remove globular clusters, dwarf galaxies, Sagittarius, and Large and Small Magellanic Clouds (LMC, SMC) to get the smooth stellar halo.
We use the same method as \citet{iorio2021MNRAS} to cut the globular clusters \citep{Harris1996J} and dwarf galaxies \citep{Mateu2023MNRAS}.
Similar to \citet{iorio2021MNRAS}, we remove all stars with $\lvert \tilde{B}-\tilde{B}_{\rm Sgr}\rvert\,\textless\,9^{\circ}$ and $\lvert \tilde{\Lambda}-\tilde{\Lambda}_{\rm Sgr}\rvert\,\textless\,50^{\circ}$ to exclude the core of Sgr dwarf, where $\tilde{B}$ and $\tilde{\Lambda}$ are the latitude and longitude in the coordinate system aligned with the Sgr stream as defined in \citet{Belokurov2014MNRAS} and $\tilde{B}_{\rm Sgr}=4.24^{\circ}$ and $\tilde{\Lambda}_{\rm Sgr}=-1.55^{\circ}$ represent the position of Sgr dwarf. 
In order to completely remove the Sgr stream, we remove the RRab stars satisfying $d_{\odot}\,\textgreater\,15\,{\rm kpc}$ and $\lvert \tilde{B} \rvert\,\textless\,11^{\circ}$ as done by \citet{Wegg2019MNRAS}.
Similar to \citet{Ye2023MNRAS}, we remove all RR Lyrae within an angular distance of $16^{\circ}$ ($12^{\circ}$) from the centre of LMC (SMC) as reported in \citet{van2016ApJ}.
To eliminate the Galactic disk and heavily extincted regions, we consider only the RRab stars with $\lvert z\rvert\,\textgreater\,6\,{\rm kpc}$ and $\lvert b\rvert\,\textgreater\,10^{\circ}$.
To remove the majority of the likely contaminants, we apply the following selection cuts: $RUWE\,\textless\,1.4$ and ${\rm phot\underline{~}bp\underline{~}rp\underline{~}excess\underline{~}factor}\,\textless\,3$.
Finally, we select stars with $\phi_{31}$ and $R_{21}$ measurements due to the high reliability of metallicity derived from them.
In Figure \ref{figRRLall}, we show the clean RRL sample.

\section{Methods}\label{Methods}

Based on our RRL sample, we apply the Bayesian approach to the Gaussian mixture model to investigate the contributions of GSE to HAC and VOD. 
Note that the RRab stars only have 2D velocity information, so they are fitted with the two-dimensional tangential projection of the three-dimensional Gaussian mixture model.
To probe the possible compositions of low-eccentricity stars, based on our K-giant sample, we apply the selection criteria of known substructures to identify their member stars in the HAC and VOD regions.
Finally, we consider a doubly broken power law (DBPL), coupled with a rotational and triaxial ellipsoid, to study the reasons for the tilt and breaks of the stellar halo.

\subsection{Gaussian mixture model}\label{Gaussian mixture model used for RRab stars}

\begin{figure}
\centering
	\includegraphics[width=\columnwidth]{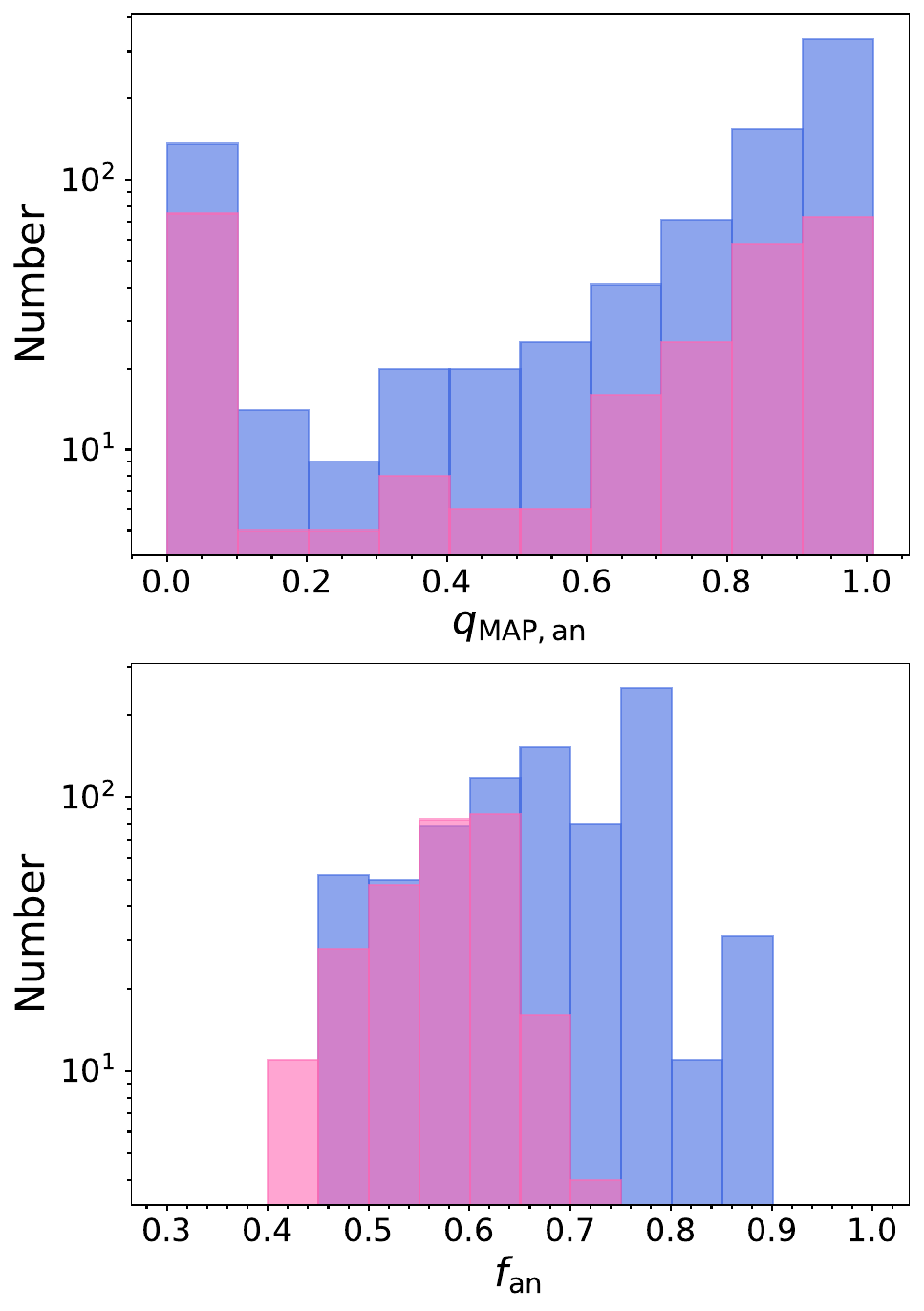}
    \caption{Top: Distribution of the maximum posterior probability of belonging to the radially anisotropic kinematic component for the RRab stars from HAC (royal blue) and VOD (hot pink). Bottom: Distribution of weights of radially anisotropic component for HAC (royal blue) and VOD (hot pink).}
    \label{fighacvod_qan_fan}
\end{figure}

The GMM we used includes two stellar halo components, i.e., the  isotropic metal-poor one and the radially anisotropic lightly metal-rich one from the GSE merger \citep{Belokurov2018MNRAS,Helmi2018Natur,Deason2018ApJ,Kruijssen2019MNRAS,Myeong2018ApJ1,Myeong2018ApJ2}.
\citet{Wegg2019MNRAS} measured the intrinsic kinematics of the inner stellar halo between 1.5\,kpc and 20\,kpc from the Galactic centre assuming a Gaussian distribution of velocities at each point in space, and then applied these kinematic measurements to the Jeans equation to infer the averaged gravitational acceleration field.
\citet{iorio2021MNRAS} added a further generalization considering the intrinsic velocities as a composition of multiple multivariate normal distributions.
In our work, we further supplement the metallicity component on the basis of \citet{iorio2021MNRAS}, which can alleviate degeneracy issues while excluding metal-rich anisotropic components such as Splash \citep{Belokurov2020MNRAS}.
The likelihood for this GMM is defined as
\begin{equation}
\begin{aligned}
\mathcal{L}\,(\textit{\textbf{D}} \lvert \Theta) & = 
\prod_{i}(f_{\rm iso}\mathcal{L}_{\rm iso}(\textit{\textbf{D}}_{i} \lvert \Theta)+f_{\rm an}\mathcal{L}_{\rm an}(\textit{\textbf{D}}_{i} \lvert \Theta))^{\frac{1}{\mathcal{S}_i}}\\
\end{aligned}\label{likelihood}
\end{equation}
where, $i$ is a product over the data points, $\textit{\textbf{D}}_{i}$ is the vector of the $i$-th data points, and $\Theta$ is the vector of model parameters,
while $f_{\rm iso}$ and $f_{\rm an}$, satisfying $f_{\rm an}+f_{\rm iso}=1$, denote the fraction contributions from the isotropic and radially anisotropic halo components, respectively.
$\mathcal{S}_i$ is the selection function of a star $i$.
\citet{Mateu2024RNAAS} have given the selection function of \textit{Gaia} DR3 RR Lyrae under the selections about $RUWE$ and ${\rm phot\underline{~}bp\underline{~}rp\underline{~}excess\underline{~}factor}$.
We also calculate the proportion of stars with $\phi_{31}$ and $R_{21}$ measurements in each volume unit divided by \citet{Mateu2024RNAAS}, because metallicity values of these stars are more reliable and they will be fitted.
Following \citet{Wu2022ApJ}, the selection function is converted to the likelihood, as shown in Equation (\ref{likelihood}), which means that the weight of each star has been adjusted accordingly.
The likelihood of the isotropic stellar halo for the $i$-th data point is given by
\begin{equation}
\begin{aligned}
\mathcal{L}_{\rm iso}(\textit{\textbf{D}}_{i} \lvert \Theta) & = 
\mathcal{N}(\textit{\textbf{v}}_{\bot i} \lvert \bar{\textbf{v}}_{\bot i}^{\rm iso},\,\textbf{C}_{\bot i}^{\rm iso})\mathcal{N}({\rm [Fe/H]}_{i} \lvert \mu_{\rm [Fe/H]}^{\rm iso}, \sigma^{\rm iso\ \ \ \ \ \ 2}_{\rm [Fe/H],i} )\\
\end{aligned}
\end{equation}
The likelihood of the radially anisotropic or GSE component is given by 
\begin{equation}
\begin{aligned}
\mathcal{L}_{\rm an}(\textit{\textbf{D}}_{i} \lvert \Theta) &= [\frac{1}{2}\mathcal{N}(\textit{\textbf{v}}_{\bot i} \lvert \bar{\textbf{v}}^{\rm an+}_{\bot i},\,\textbf{C}_{\bot i}^{\rm an})+\frac{1}{2}\mathcal{N}(\textit{\textbf{v}}_{\bot i} \lvert \bar{\textbf{v}}^{\rm an-}_{\bot i},\,\textbf{C}_{\bot i}^{\rm an})]\\
& \times\mathcal{N}({\rm [Fe/H]}_{i} \lvert \mu_{\rm [Fe/H]}^{\rm an}, \sigma^{\rm an\ \ \ \ \ \ 2}_{\rm [Fe/H],i} ),
\end{aligned}
\end{equation}

\begin{equation}
\begin{aligned}
\boldsymbol{v}_{\bot i} &= 
\begin{bmatrix}
    v_{b,i}\\
    v_{l,i}\\
\end{bmatrix}\\
\bar{\textbf{v}}_{\bot i} &= \textbf{R}_{\bot i}\,\bar{\textbf{v}}=\textbf{R}_{\bot i}\,
\begin{bmatrix}
    \bar{\text{v}}_r\\
    \bar{\text{v}}_{\theta}\\
    \bar{\text{v}}_{\phi}
\end{bmatrix}\\
\textbf{R}_{i} &= \textbf{U}_{1,i}\,\textbf{U}_{{\rm s},i}^{-1}\\
\textbf{C}_{\bot i} &= \textbf{S}_{\bot i} + \boldsymbol{\Lambda}_{\bot i}=\textbf{S}_{\bot i} + \textbf{R}_{\bot i}\boldsymbol{\Sigma}\textbf{R}_{\bot i}^{\boldsymbol{\top}}\\
\boldsymbol{\Lambda}_{i} &= \textbf{R}_i\boldsymbol{\Sigma}\textbf{R}_i^{\boldsymbol{\top}}= 
\begin{bmatrix}
\Lambda_{bb} & \Lambda_{bl} & \Lambda_{b\,{\rm los}}\\
\Lambda_{lb} & \Lambda_{ll} & \Lambda_{l\,{\rm los}}\\
\Lambda_{{\rm los}\,b} & \Lambda_{{\rm los}\,l} & \Lambda_{{\rm los}\,{\rm los}}
\end{bmatrix},
\end{aligned}
\end{equation}
where $\textbf{R}_{\bot i}$ is the rotation matrix $\textbf{R}_i$ without the line-of-sight direction, and it is a $2\times3$ matrix, while the intrinsic velocity dispersion $\boldsymbol{\Lambda}_{\bot i}$ is a $2\times2$ matrix, that is, $\boldsymbol{\Lambda}_i$ without the line-of-sight direction.
$\boldsymbol{\Sigma}$ is the velocity dispersion tensor in spherical coordinates.
The $\textbf{S}_{\bot i}$ represents the measurement covariance for 2D velocity. 
Note that we do not add the `iso' and `an' superscripts for the sake of clarity.
For the radially anisotropic component, the means $\bar{\textbf{v}}^{\rm an+}_{\bot i}$ and $\bar{\textbf{v}}^{\rm an-}_{\bot i}$ correspond to two distinct lobes at high positive and high negative radial velocity.
For the kinematic parameters we used the priors from \citet{iorio2021MNRAS} except $\mathcal{N}\,(0,\,300)$ for $\bar{v}_{r}^{\rm an}$, because \citet{Wu2022ApJ} recently have shown that it is significantly larger than zero.
\citet{Wang2022MNRAS} identify substructures of the Galactic halo using 3,\,003 RRab stars and found a mean of $\rm [Fe/H]=-1.56$ with $\sigma_{\rm [Fe/H]}=0.31$ for GSE.
Therefore, we set reasonable priors for the metallicity parameters, i.e. $0\,\textless\,\sigma_{\rm [Fe/H]}^{\rm iso}\,\textless\,1$, $0\,\textless\,\sigma_{\rm [Fe/H]}^{\rm an}\,\textless\,1$ and $-1.8\,\textless\,\mu_{\rm [Fe/H]}^{\rm an}\,\textless\,-1$.
We separate the RRL sample into 62 cylindrical $R$, $\lvert z\rvert$ bins with an average Poisson signal-to-noise ratio of 15 using \textbf{\tt\string vorbin} python package \citep{vorbin2003MNRAS}. In each of these bins, the intrinsic distribution of velocities is considered constant.
We sample the posterior using the affine-invariant ensemble sampler Markov Chain Monte Carlo \citep[MCMC,][]{goodman2010ensemble} method implemented in the \textbf{\tt\string emcee} python package \citep{Foreman2013PASP}.

\subsection{Identification of substructures in the HAC and VOD}\label{Identification of substructures in the HAC and VOD}

\begin{figure*}
\centering
	\includegraphics[width=0.99\textwidth]{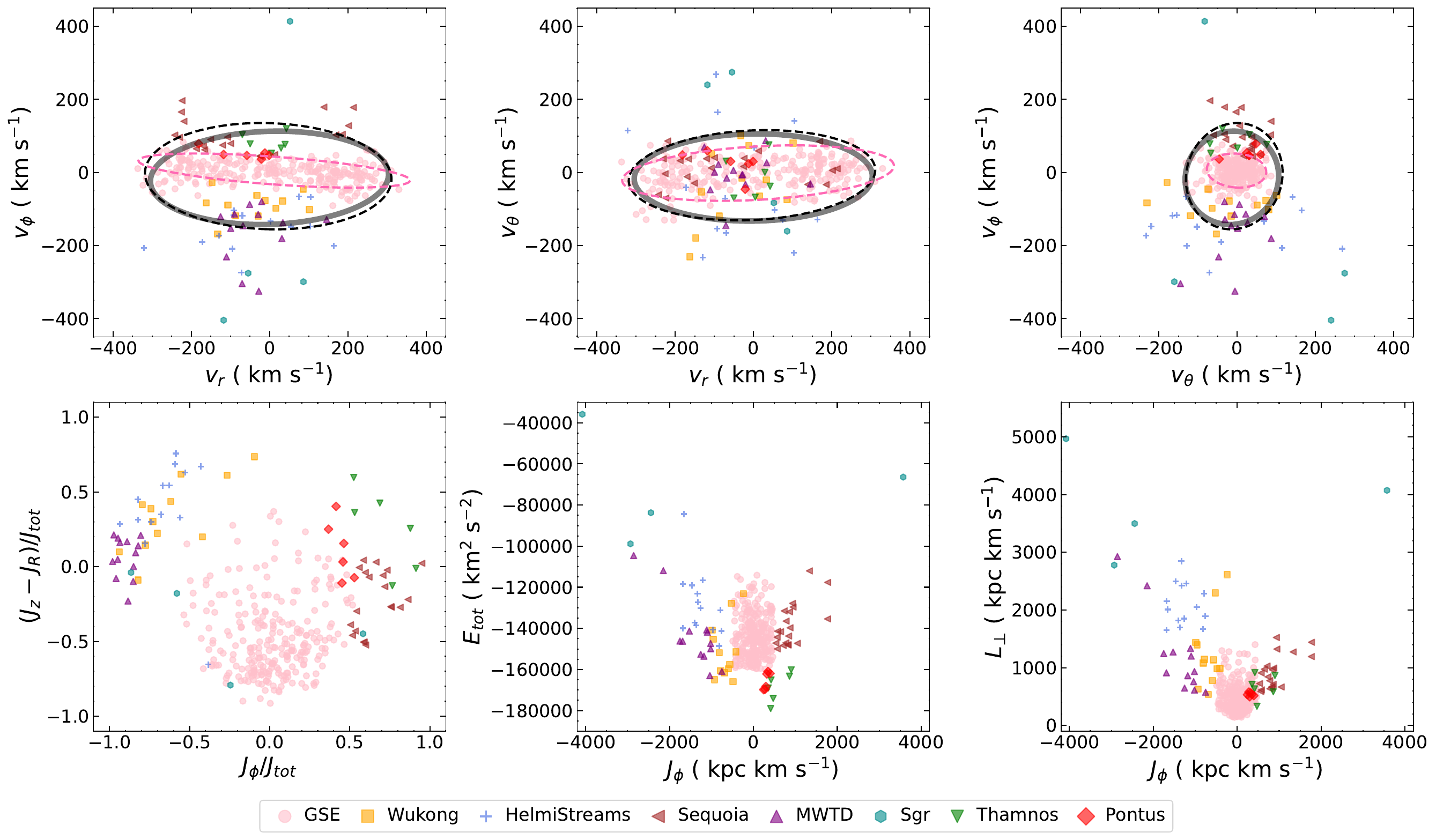}
    \caption{Kinematics and dynamical maps of substructures identified in HAC. Top: velocity distribution in spherical coordinates.
    The dashed velocity ellipsoids represent the Gaussian component fits to the HAC sample (black) and the GSE stars (pink) in the HAC sample, and the black solid velocity ellipsoid marks the Gaussian component fit to the combination of the HAC and VOD samples.
    Note that markers for the eight substructures mentioned in Section \ref{Identification of substructures in the HAC and VOD} are given in the legend. Bottom left: substructures distributed in the projected action space, i.e. $J_{\phi}/J_{\rm tot}$ vs. $(J_z-J_R)/J_{\rm tot}$, where $J_{\rm tot}=\sqrt{J_R^2+J_{\phi}^2+J_{z}^2}$. Bottom center: energy $E_{\rm tot}$ vs. the azimuthal component of the action (i.e., $J_{\phi}\equiv L_z$). Bottom right: $J_{\phi}$ vs. $L_{\bot}$, where the orthogonal component of the angular momentmum $L_{\bot}=\sqrt{L_x^2+L_y^2}$. }
    \label{figallmergers_hac}
\end{figure*}
\begin{figure*}
\centering
	\includegraphics[width=0.99\textwidth]{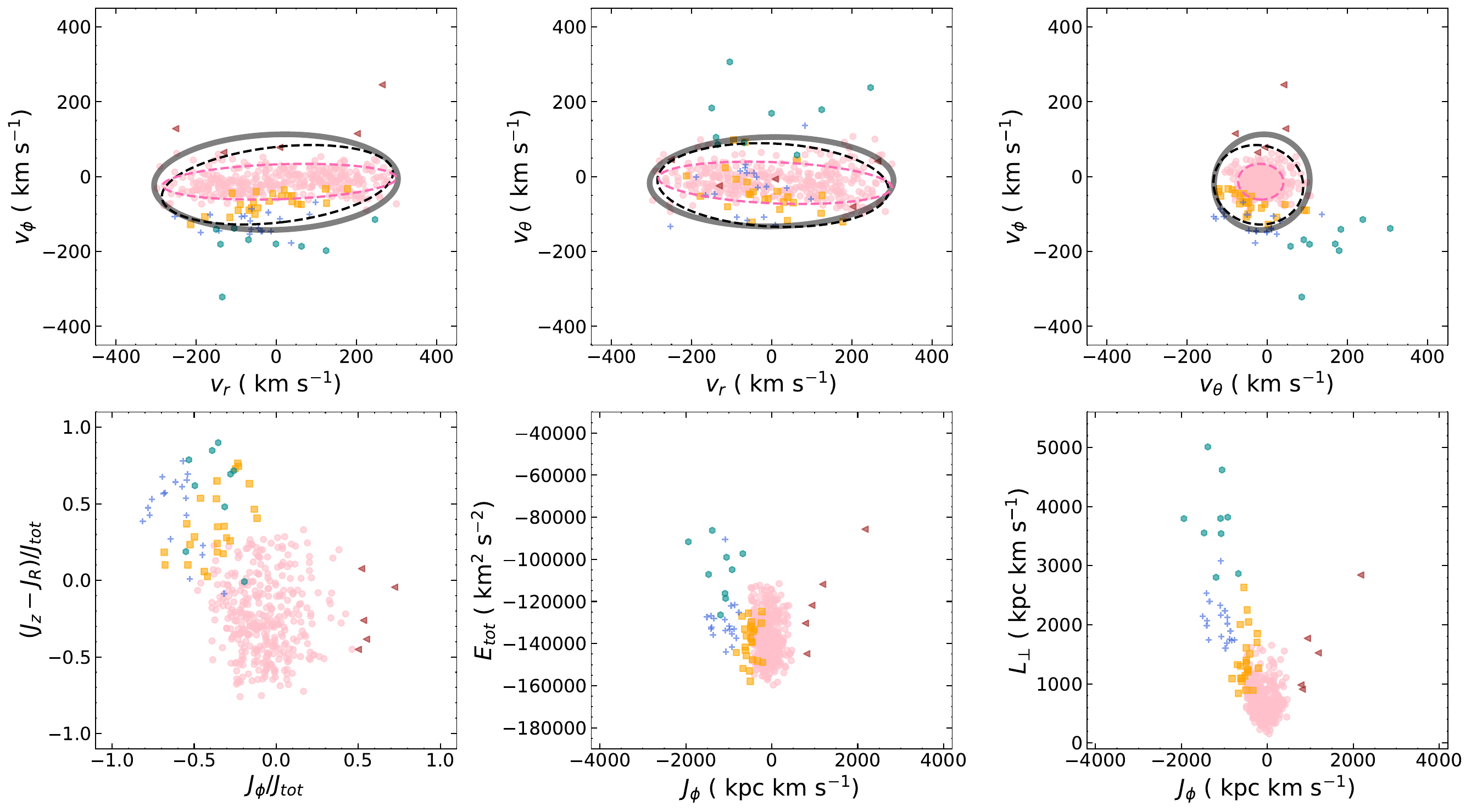}
    \caption{Substructures from VOD in velocity and dynamical spaces. 
    Note that the dashed lines represent the velocity ellipsoids of the VOD (black) sample and its GSE (pink) member stars and the solid line still indicates the velocity ellipsoid of a total of the HAC and VOD samples. 
    The markers for distinguishing substructures are identical to those in Figure \ref{figallmergers_hac}.}
    \label{figallmergers_vod}
\end{figure*}

We attempt to identify those previously known substructures from K giants of HAC and VOD, and list the selection criteria for eight well-known substructures, as
\begin{itemize}
    \item Sagittarius: $L_y\,\textgreater\,0.3J_{\phi}+2.5\times10^3{\rm\,kpc\,km\,s^{-1}}$ \citep{Naidu2020ApJ}
    \item Gaia-Sausage-Enceladus: $(\lvert J_{\phi}\rvert\,\textless\,500{\rm\,kpc\,km\,s^{-1}})\,\wedge\,(-1.6\,\textless\,E/[10^5\,{\rm km^2\,s^{-2}}]\,\textless\,-1.1)\,\wedge\,(e\,\textgreater\,0.7)\,\wedge\\({\rm excluding\ all\ previously\ defined\ substructures})$ \citep{Naidu2020ApJ,Horta2023MNRAS}
    \item Helmi Streams: $(-1.7\,\textless\,J_{\phi}/[{\rm 10^3\,kpc\,km\,s^{-1}}]\,\textless\,-0.75)\,\wedge\,(1.6\,\textless\,L_{\bot}/[{\rm 10^3\,kpc\,km\,s^{-1}}]\,\textless\,3.2)\,\wedge\\({\rm excluding\ all\ previously\ defined\ substructures})$ \citep{Naidu2020ApJ}
    \item Sequoia: $(E/[10^5\,{\rm km^2\,s^{-2}}]\,\textgreater\,-1.5)\,\wedge\,(J_{\phi}/J_{\rm tot}\,\textgreater\,0.5)\,\wedge\,((J_z-J_R)/J_{\rm tot}\,\textless\,0.1)\,\wedge\\({\rm excluding\ all\ previously\ defined\ substructures})$ \citep{Myeong2019MNRAS}
    \item Thamnos: $(-1.8\,\textless\,E/[10^5\,{\rm km^2\,s^{-2}}]\,\textless-1.6)\,\wedge\,(0.2\,\textless\,J_{\phi}/[{\rm 10^3\,kpc\,km\,s^{-1}}]\,\textless1.5)\,\wedge\,({\rm [Fe/H]}\,\textless\,-1.6)\,\wedge\,(e\,\textless\,0.7)\,\wedge\,({\rm excluding\ all\ previously\ defined\ substructures})$ \citep{Naidu2020ApJ,Horta2023MNRAS}
    \item Pontus: $(0.5\,\textless\,e\,\textless\,0.8)\, \wedge\, (0.245\,\textless\,J_R/[{\rm 10^3\,kpc\,km\,s^{-1}}]\,\textless\,0.725)\, \wedge\, (-0.005\,\textless\,J_{\phi}/[{\rm 10^3\,kpc\,km\,s^{-1}}]\,\textless\,0.470)\, \wedge \, (0.115\,\textless\,J_z/[{\rm 10^3\,kpc\,km\,s^{-1}}]\,\textless\,0.545)\, \wedge \, (-1.72\,\textless\,E/[{\rm 10^5\,km^2\,s^{-2}}]\,\textless\,-1.56)\,\wedge\,(8\,{\rm kpc}\,\textless\,r_{\rm apo}\,\textless\,13\,{\rm kpc})\,\wedge\,(1\,{\rm kpc}\,\textless\,r_{\rm peri}\,\textless\,3\,{\rm kpc})\,\wedge\,(0.390\,\textless\,L_{\bot}/[{\rm 10^3\,kpc\,km\,s^{-1}}]\,\textless\,0.865)\,\wedge\,({\rm excluding\ all\ previously\ defined\ substructures})$ \citep{Malhan2022ApJ}
    \item Wukong: $(-1\,\textless\,J_{\phi}/[{\rm 10^3\,kpc\,km\,s^{-1}}]\,\textless\,-0.2)\,\wedge\,(-1.7\,\textless\,E/[{\rm 10^5\,km^2\,s^{-2}}]\,\textless\,-1.2)\,\wedge\,(${\rm [Fe/H]}\,\textless\,-1.45$)\,\wedge\,(0.4\,\textless\,e\,\textless\,0.7)\,\wedge\,(\lvert z\rvert\,\textgreater\,3\,{\rm kpc})\,\wedge\,({\rm excluding\ all\ previously\ defined\ substructures})$ \citep{Naidu2020ApJ,Horta2023MNRAS}
    \item Metal-weak Thick Disk: $(-2.5\,\textless\,{\rm[Fe/H]}\,\textless\,-0.8)\,\wedge\,(J_{\phi}/J_{\rm tot}\,\textless\,-0.8)\,\wedge\,(\lvert J_z-J_R\rvert/J_{\rm tot}\,\textless\,0.3)\,\wedge\\({\rm excluding\ all\ previously\ defined\ substructures})$ \citep{Naidu2020ApJ}
\end{itemize}
The selection criteria for Sgr mentioned above are completely consistent with those of \citet{Naidu2020ApJ}, because they have verified that the criteria can select $\textgreater\,99.5\%$ of the star particles in the model of \citet{Law2010ApJ}.
Note that the \textit{x,\ y}-axes here are opposite those of \citet{Naidu2020ApJ} respectively.
To extract high-purity GSE from HAC and VOD, its selection is a combination of \citet{Naidu2020ApJ} and \citet{Horta2023MNRAS}.
\citet{Horta2023MNRAS} also applied the \citet{McMillan2017MNRAS} Galactic potential to determine energy, so there is no energy offset.
Thus, for Thamnos, the energy cuts are from \citet{Horta2023MNRAS}, and the selection criteria except for those involving energy come from \citet{Naidu2020ApJ}.
Following the previous studies \citep{Koppelman2019A&A,Naidu2020ApJ,Horta2023MNRAS}, we rely on the $L_z-L_{\bot}$ selection to determine the Helmi Streams.
The selection criteria for Sequoia and Pontus were taken from \citet{Myeong2019MNRAS} and \citet{Malhan2022ApJ}.
For Wukong, we adopt the strict selection criteria of \citet{Horta2023MNRAS} without energy offset. 
Note that in their work the \textit{y}-axis points towards the Galactic rotation.
Since the K giant sample does not contain very detailed chemical information, we strengthen the dynamical constraints for less contaminants, referring to the MWTD stars clumped in the $(J_z-J_R)/J_{\rm tot}$ vs. $J_{\phi}/J_{\rm tot}$ space (Figure 16 of \citet{Naidu2020ApJ}).

\subsection{Density Fitting}\label{Density Fitting}
\begin{figure}
	\includegraphics[width=\columnwidth]{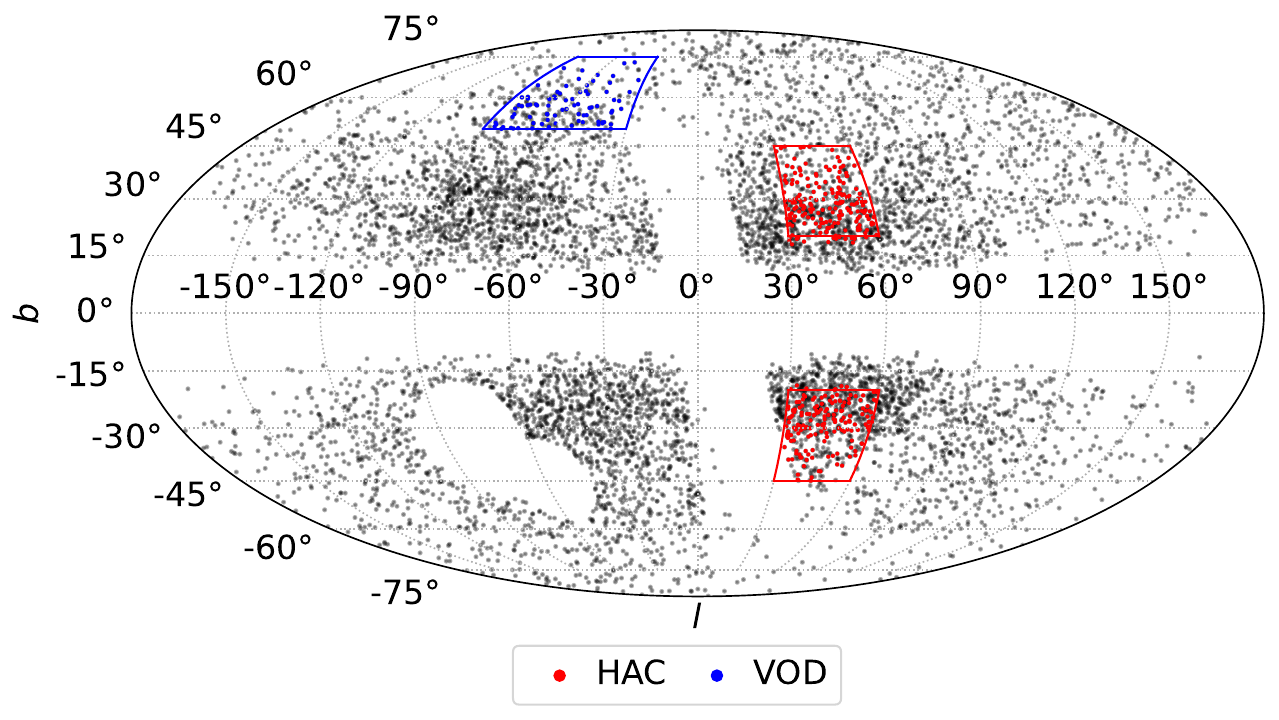}
    \caption{Spatial projection in Galactic coordinates of all stars (black) with $q_{\rm MAP,an}\,\textgreater\,0.9$, red and blue dots indicate HAC and VOD stars satisfying this condition.}
    \label{figanisotropic_halo}
\end{figure}

\begin{figure*}
	\includegraphics[width=\textwidth]{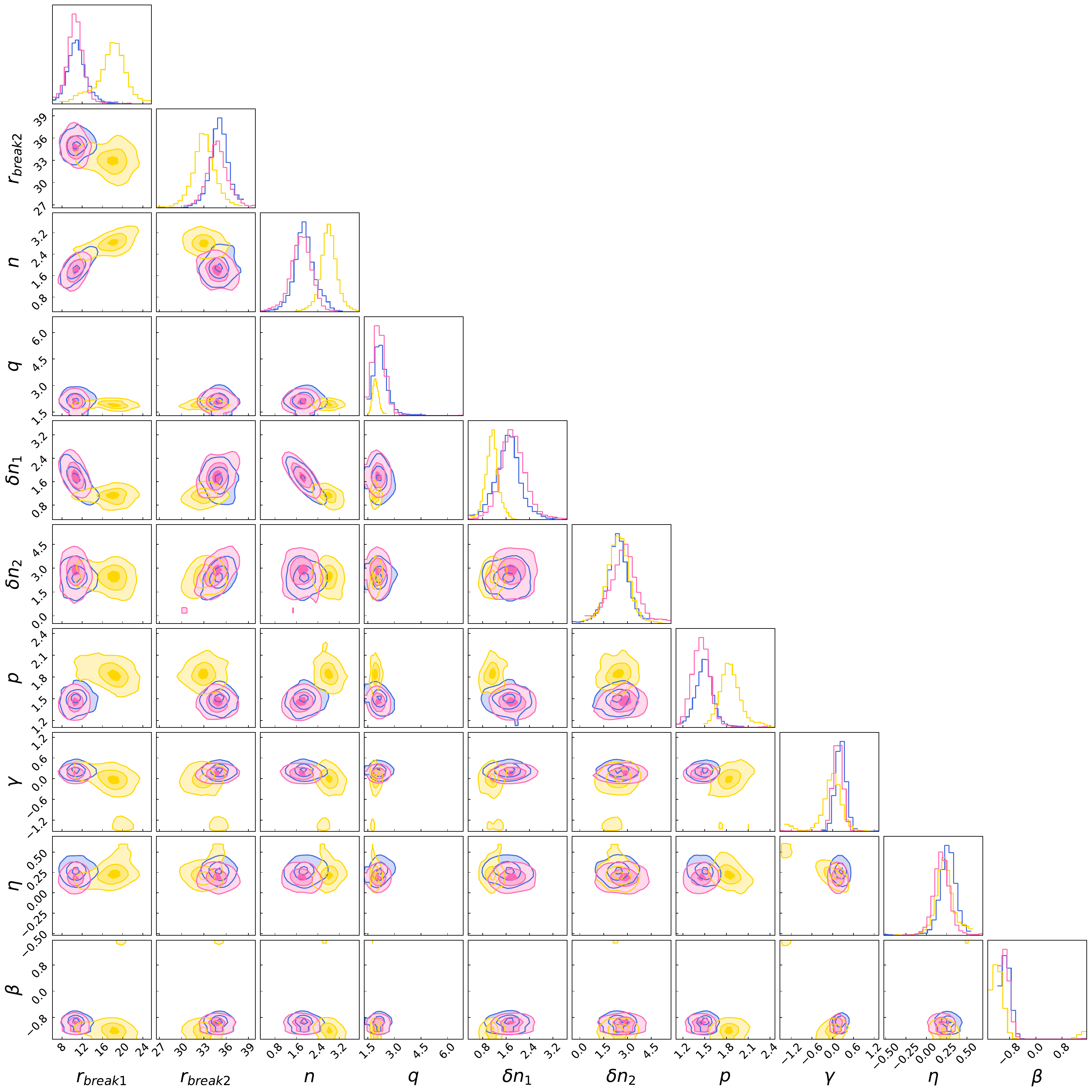}
    \caption{One- and two-dimensional projections of the posterior probability distributions of parameters of the doubly broken power law fitting results, within $r\,\textless\,40\,{\rm kpc}$, for three samples, including the clean RRL sample (royal blue), the sample with $q_{\rm MAP,an}\,\textgreater\,0.9$ (yellow) and the sample with the HAC and VOD removed (hot pink).}
    \label{figdensityfitting}
\end{figure*}

Following \citet{Ye2023MNRAS}, the broken power-law model is given by
\begin{equation}
\rho_{\rm halo} = \rho_{\odot}^{\rm RRab}\left( \frac{R_{\odot}}{r_{\mathrm{e}}} \right)^{\left (n + \sum\limits_{i=0}^{m} \left(0.5\delta n_{i} +\left( \frac{\delta n_{i}}{\pi} \right){\rm arctan}\left((10^{a_{i}})\times\left(r_{\mathrm{e}}-r_{\mathrm{break},i}\right)\right)\right)   \right) }. 
\end{equation}
where $a_i$ is a parameter associated with the break scale, that is, a larger break scale corresponds to a smaller $a_i$.
Therefore, it can reflect the characteristics of the structure making broken power law, for example, a compact or diffuse structure corresponds to a small or large break scale (i.e., large or small $a_i$).
Here we set $a_i$ to 3 to make the model very close to the general broken power law \citep[e.g.,][]{Xue2015ApJ,Hernitschek2018ApJ,Iorio2018MNRAS}, in order to facilitate comparison with previous results.
$m$ is the number of breaks in the density profile.
$r_{\rm e}$ represents elliptical radius, namely $r_{\rm e}=\sqrt{\hat{x}^{2}+(\hat{y}/p)^{2}+\left(\hat{z}/q\right)^{2}}$, here $ p $ and $ q $ are, respectively, the $y$-to-$x$ and $z$-to-$x$ ellipsoid axial ratios ($\hat{\textbf{r}} = \textbf{R}_{y}(\beta)\textbf{R}_{x}(\eta)\textbf{R}_{z}(\gamma)\textbf{r}$, $\textbf{R}_{y}(\beta)$, $\textbf{R}_{x}(\eta)$ and $\textbf{R}_{z}(\gamma)$ are clockwise rotation matrices around the $y\text{-axis, }x\text{-axis and }z\text{-axis}$, respectively, so our rotation transformation aims to first stabilize the direction of the major axis with $\gamma$ and $\eta$, and then determine those of the minor axes.). 
\textit{n} is the power-law index or slope that describes the shape of the density profile, and a large value of $\delta n$ means a prominent slope variation at $r_{\rm break}$, that is, a significantly broken density profile.
The likelihood is normalized as follows:
\begin{equation}
\begin{aligned}
{\rm ln}\,\mathcal{L}
&=\sum_{i}{\rm ln}\,\mathcal{L}(r_{i},\theta_{i},\phi_{i}\lvert \Theta_{\rm d})\\
&=\sum_{i}{\rm ln}\,\frac{ \rho_{\rm halo}(r_{i},\theta_{i},\phi_{i}\lvert\,\Theta_{\rm d})\mathcal{S}(r_{i},\theta_{i},\phi_{i})\lvert \textbf{\textit{J}}_i \rvert}{\iiint_V \rho_{\rm halo}(r,\theta,\phi\lvert\,\Theta_{\rm d})\mathcal{S}(r,\theta,\phi)\lvert \textbf{\textit{J}} \rvert \,dr\,d\theta\,d\phi}, 
\end{aligned}
\end{equation}
where the Jacobian term $\lvert J\rvert=r^2\,\text{sin}\,\theta$ reflects the transformation from $(x,\,y,\,z)$ to $(r,\,\theta,\,\phi)$. Note that the normalization integral is over the volume of $r\,\textless\,40\,{\rm kpc}$.
$\mathcal{S}$ represents the product of selection functions for all structures or areas we spatially removed, including the regions of $\lvert z \rvert\,\textless\,6\,{\rm kpc}$ and $\lvert b\rvert\,\textless\,10^{\circ}$, globular clusters, dwarf satellites, LMC, SMC, the Sgr dwarf and stream. 
Therefore, the selection functions in these regions are all set to zero.
We also consider the selection effects of Fourier parameter measurements as well as the \textit{Gaia} DR3 RR Lyrae under the selection cuts about $RUWE$ and ${\rm phot\underline{~}bp\underline{~}rp\underline{~}excess\underline{~}factor}$ \citep[see][for details]{Mateu2024RNAAS}.
Otherwise, selection effects due to preparing the clean RRL sample and magnitude-limited observations will affect fitting results.
Here, a very conservative prior is adopted on the basis of previous studies \citep[e.g.,][]{Hernitschek2018ApJ,Iorio2018MNRAS,Han2022AJ} as follows: $a_i\in\delta(3)$,\ $\delta n_i \in (0.1,6)$,\ $n\in(0.1,10),\ q\in(0.1,10),\ r_{\rm break,1}\in(6,26),\ r_{\rm break,2}\in(26,40),\ p\in(0.1,10),\ \beta,\ \eta\ {\rm and}\ \lambda\in(-0.5\pi,0.5\pi)$.

\section{Results}\label{Results}

\subsection{Contribution of the Gaia-Sausage-Enceladus to the HAC and VOD}\label{Contribution of the Gaia-Sausage-Enceladus to the HAC and VOD clouds}

\citet{Simion2019MNRAS} and \citet{Yan2023A&A} suggested a common origin of the HAC and VOD by comparing their kinematic properties, chemical abundances, and orbits.
\citet{Boubert2019MNRAS} found that the VOD is aligned with the Magellanic Stream, and suggested that it is either debris from a disrupted dwarf galaxy as a member of the Vast Polar Structure or SMC debris from a tidal interaction between the SMC and LMC 3 Gyr ago.
\citet{Perottoni2022ApJ} suggested that the HAC and VOD originate from the GSE merger based on the eccentricity and elemental abundances analysis. 
In this work, we investigate their kinematics to verify that they are dominated by the radially anisotropic stellar halo, namely GSE.

We adopt the stochastic variable ($q_{\rm an}$) defined by \citet{iorio2021MNRAS} to describe the probability of a star belonging to the GSE,
\begin{equation}
\begin{aligned}
q_{\rm an} & = \frac{f_{\rm an}\mathcal{L}_{\rm an}}{f_{\rm an}\mathcal{L}_{\rm an}+f_{\rm iso}\mathcal{L}_{\rm iso}}.
\end{aligned}
\end{equation}
Figure \ref{fighacvod_qan_fan} shows the distributions of stochastic variable ($q_{\rm MAP, an}$) and radially anisotropic weights for RRab stars of HAC (royal blue) and VOD (hot pink).
Note that $q_{\rm MAP, an}$ represents an object of the maximum posterior probability belonging to the GSE, and a $q_{\rm MAP, an}$ closer to 1 means a larger probability of belonging to the GSE.
In either of HAC and VOD, most of stars satisfying $q_{\rm MAP,an}\,\textgreater\,0.6$ presumably belong to the GSE debris, a few stars with $q_{\rm MAP,an}\,\textless\,0.2$ are more likely to be isotropic, and the remnants may be indistinguishable mixtures of the two components.
The weight of radial anisotropy is greater than 0.5 for most stars from HAC and VOD, and the medians of weight for HAC and VOD are $0.67^{+0.09}_{-0.07}$ and $0.57^{+0.07}_{-0.06}$, respectively.
The mean metallicities of the anisotropic stellar halo in the HAC and VOD fields are $\mu_{\rm [Fe/H]}^{\rm an}=1.57^{+0.02}_{-0.03}$ and $1.58^{+0.02}_{-0.01}$, respectively, which are consistent with that of GSE detected by \citet[][]{Wang2022MNRAS} using RRab stars.
Combined with the substantial overlap in various chemical-abundance spaces among HAC, VOD, and GSE \citep{Perottoni2022ApJ} and the above results, we suggest that the radially accreted GSE is likely to be the major contribution of the HAC and VOD.

\subsection{Analysis of substructures}\label{Analysis of substructures}

Here we analyze the dynamical substructures subdivided in Section \ref{Identification of substructures in the HAC and VOD} using the K giant samples of HAC and VOD.
Of 568 and 544 K giants from HAC and VOD, we respectively identified 326 and 402 stars as members of substructures under our selection criteria.
Figures \ref{figallmergers_hac} and \ref{figallmergers_vod} show the substructures from HAC and VOD in the Galactocentric spherical velocities and dynamical spaces.
In order to illustrate the shape of the velocity distributions, we model the data as a single-component multivariate Gaussian using the Extreme Deconvolution method \citep{Bovy2011AnApS} as implemented in \textbf{\tt\string astroML} \citep{vanderplas2012introduction}.
As expected, the velocity ellipsoid in the HAC or VOD sample (black dashed line) resembles that of HAC and VOD as a whole (solid line) due to a common origin for them.
The velocity ellipsoid of GSE selected from HAC or VOD (pink dashed line) is extremely oblate vertically or prolate radially, and approximately centres on that of HAC (VOD).
The velocity ellipsoid shape can be summarized using the anisotropy parameter $\beta_{\rm an} = 1-(\sigma_{\theta}^2+\sigma_{\phi}^2)/2\sigma_{r}^2$ \citep[][]{binney2008st}, and the subscript `an' is here to avoid confusion with the $\beta$ rotational angle around \textit{y}-axis in Selection \ref{Density Fitting}.
The orbital anisotropy parameters of HAC and VOD, namely $\beta_{\rm an}\sim0.816$ and 0.857, are very close to those of individual GSE (i.e., $\beta_{\rm an}\sim0.967$, 0.968), and it means that the shape of the velocity ellipsoid of HAC (VOD) is flat vertically and extended radially as that of the GSE in HAC (VOD). 
Therefore, this kinematically supports the dominance of the GSE stars for both.
The GSE selected from VOD is more prograde and has higher energy than that from HAC ($v_{\phi}=-13.26\,{\rm km\,s^{-1}}$ and $E=-136,\,133\,{\rm km^2\,s^{-2}}$ for VOD, $v_{\phi}=1.19\,{\rm km\,s^{-1}}$ and $E=-143,\,320\,{\rm km^2\,s^{-2}}$ for HAC), following the fact that the GSE debris at high energies tends to be more prograde \citep{Belokurov2023MNRAS,Ye2024}.
In the projected action space, comparing the GSE debris of HAC with that of VOD, we find that the GSE stars in HAC tend to concentrate in the region of $(J_z-J_R)/J_{\rm tot}\,\textless\,-0.5$, while for VOD, the GSE stars are mostly distributed in the region of $(J_z-J_R)/J_{\rm tot}\,\textgreater\,-0.5$ and tend to satisfy $J_{\phi}/J_{\rm tot}\,\textless\,0$, indicating that the GSE stars in VOD move on relatively polar orbits.

Except for the major GSE merger, a lot of retrograde stars from the other substructures, including Pontus, Thamnos and Sequoia, exist in the K giant sample of HAC, while only five retrograde stars defined Sequoia are detected in the VOD sample.
The apocentric distances of Pontus stars range within $8\text{--}13\,{\rm kpc}$ \citep{Malhan2022ApJ}, and here 21\% of the K giant sample of VOD satisfy $r\,\textless\,13\,{\rm kpc}$, which may make it difficult to be found in the VOD sample.
Most of the Thamnos stars selected by \citet{Naidu2020ApJ} using the H3 data are within 10\,kpc from the Galactic centre (see Figure 13 in original paper), and here we adopt selection criteria from \citet{Naidu2020ApJ} and \citet{Horta2023MNRAS}, so there are no Thamnos stars in the K giant sample of VOD ($\textgreater99\%$ with $r\,\textgreater\,10\,{\rm kpc}$). 
Note that we do not rule out other possible reasons, for example, Pontus, Thamnos and Sequoia share a common progenitor galaxy with the GSE. 
\citet{Amarante2022ApJ} suggest that Pontus is most likely part of the GSE merger because Pontus are wrapped by GSE in energy and action space and has the same metallicity as GSE.
In this work, by comparing substructures in HAC and VOD, we note that the reduction behaviour of retrograde component of GSE is similar to the decay behaviour of number of member stars of retrograde Pontus, Thamnos and Sequoia.
Although the metallicities of Thamnos and Sequoia are slightly different from that of GSE, it is still likely that a common progenitor galaxy has unevenly distributed metallicities, that is, several parts with significantly distinct spatial locations have different metallicity.
We look forward to further investigating whether they originate from a single merger, but this is beyond the scope of this work.

The MWTD may be a major component for $\lvert z\rvert\,\textless\,3\,{\rm kpc}$ while a minor component at larger distances \citep{Carollo2019ApJ}, so it is not recognized in the high-latitude VOD.
However, more members of prograde ($J_{\phi}\,\textless\,0$) substructures with polar orbits (i.e.,$(J_z-J_R)/J_{\rm tot}\,\textgreater\,0$), including Sgr, Wukong and Helmi Streams, are identified in the high-latitude VOD field, as shown in the bottom-left panels of Figures \ref{figallmergers_hac} and \ref{figallmergers_vod}.
The position of the Sgr dwarf galaxy and the plane of the Sgr stream are completely missed by the HAC region, resulting in a few Sgr stars in the HAC field.
However, compared with HAC, the VOD position relatively aligned with the Sgr orbital plane may lead to the intervention of more Sgr stars.
In the projected action space, the weights of azimuthal actions of member stars of Wukong in VOD ($J_{\phi}/J_{\rm tot}\,\textgreater\,-0.5$) are generally lower than those in HAC ($J_{\phi}/J_{\rm tot}\,\textless\,-0.5$). This is contrary to the behaviour of the GSE which tends to be more prograde, perhaps suggesting that Wukong originated from a distinct progenitor galaxy that accreted with the Milky Way in a direction relatively perpendicular to the disc. 
However, the retrograde substructures show the same behaviour as GSE, that is, they become fewer or disappear.
\citet{Ye2024} found that the slopes of Wukong and Helmi Stream in $v_{\phi}-e$ space are very inconsistent with the V-shaped structure of the corresponding energy GSE. 
In comparison, MWTD, Sequoia, I'itoi+Sequoia+Arjuna, Pontus, and Thamnos are more similar, especially Thamnos, which can properly connect the V-shaped structure of GSE.
Despite the kinematic differences between them, these common features between them and GSE indicate that they are likely to originate from a common progenitor galaxy.

Finally, we provide a brief overview of all substructures and explain why other known substructures were not selected.
The Sgr stars move along highly polar orbits, such as the prominent orthogonal angular momentum and zenithal velocity, as shown in Figures \ref{figallmergers_hac} and \ref{figallmergers_vod}.
Helmi Streams and Wukong share same area in the $J_{\phi}/J_{\rm tot}$ vs. $(J_z-J_R)/J_{\rm tot}$ space, that is, $(J_z-J_R)/J_{\rm tot}\,\textgreater\,0$ and $J_{\phi}/J_{\rm tot}\,\textless\,0$, indicating that they have very polar orbits.
MWTD has larger $J_{\phi}$ and $v_{\phi}$ than those of Wukong but smaller $L_{\bot}$ and $E$, indicating that the MWTD stars move along relatively in-plane orbits. 
Here, we find some features of the GSE: radially stretched and tangentially flattened velocity ellipsoid, high anisotropy parameter, high eccentricity, and unique metallicity.
Among three retrograde substructures, namely Pontus, Thamnos, and Sequoia, apart from the difference in the ($E,\, L_z$) distribution, Pontus has a more radial high-eccentricity ($e=0.72\pm0.04$) orbit with a lower the azimuthal component of action $J_{\phi}$, and the energies of Thamnos and Pontus are significantly lower than that of Sequoia.
The aforementioned substructures are widely distributed in the inner halo, while other known substructures, such as Cetus, Aleph, Icarus and Heracles, might be situated at large distances or in the Galactic disk or centre region, so they are almost not included in HAC and VOD.

\subsection{Impacts of HAC and VOD as well as radially anisotropic halo on the shape of stellar halo}\label{Impacts of HAC and VOD as well as radially anisotropic halo on the shape of stellar halo}

\begin{table*}
\centering
\scalebox{0.8}{
\begin{threeparttable}
    \caption{\centering Fitting results.}
    \label{tabledensityfitting}
	\begin{tabular}{lcccccccccr} % four columns, alignment for each
		\hline
		Sample & $r_{\rm break1}~({\rm kpc})$ & $r_{\rm break2}~({\rm kpc})$ & $n$ & $q$ & $\delta n_1$ & $\delta n_2$ & $p$ & $\gamma$ (rad) & $\eta$ (rad) & $\beta$ (rad) \\
		\hline
         \text{Clean Sample} & $10.84^{+1.81}_{-1.39}$ & $35.06^{+0.95}_{-0.94}$ & $1.88^{+0.38}_{-0.28}$ & $2.17^{+0.40}_{-0.19}$ & $1.68^{+0.32}_{-0.42}$ & $2.37^{+0.64}_{-0.56}$ & $1.50^{+0.10}_{-0.12}$ & $0.23^{+0.07}_{-0.06}$ & $0.27^{+0.07}_{-0.07}$ & $-0.90^{+0.04}_{-0.05}$ \\
         $q_{\rm MAP, an}\,\textgreater\,0.9$&
         $18.08^{+2.04}_{-3.22}$&
         $33.03^{+1.30}_{-1.21}$&
         $2.82^{+0.24}_{-0.25}$&
         $1.93^{+0.18}_{-0.15}$&
         $1.13^{+0.18}_{-0.21}$&
         $2.43^{+0.57}_{-0.58}$&
         $1.85^{+0.14}_{-0.11}$&
         $-0.01^{+0.18}_{-0.21}$&
         $0.21^{+0.10}_{-0.07}$&
         $-1.17^{+0.08}_{-0.08}$\\
         Removal of HAC and VOD&
         $10.57^{+1.26}_{-1.19}$&
         $34.78^{+1.29}_{-1.14}$&
         $1.79^{+0.33}_{-0.37}$&
         $2.10^{+0.17}_{-0.19}$&
         $1.80^{+0.42}_{-0.33}$&
         $2.81^{+0.68}_{-0.88}$&
         $1.45^{+0.10}_{-0.11}$&
         $0.15^{+0.06}_{-0.07}$&
         $0.19^{+0.08}_{-0.08}$&
         $-0.93^{+0.05}_{-0.06}$\\
		\hline
	\end{tabular}
\end{threeparttable}}
\end{table*}

The HAC and VOD have been interpreted as apocentric pile-ups \citep{Deason2018ApJ,Naidu2020ApJ}, and they are aligned with the semimajor axis of the inner stellar halo \citep{Iorio2019MNRAS,naidu2021ApJ,Han2022ApJ}.
In Figure \ref{figanisotropic_halo}, we show the RRab stars that are most likely to belong to the radially anisotropic halo.
Figure \ref{figdensityfitting} shows that the one- and two-dimensional projections of the posterior probability distribution for the DBPL model fitting results of the three data sets within $r\,\textless\,40\,{\rm kpc}$, which are summarized in Table \ref{tabledensityfitting}.
The break radius does not show any obvious variation after the removal of HAC and VOD, but the orientation of the triaxial ellipsoid is affected.
Failure to remove difficult-to-identify HAC members at low latitudes ($\lvert b \rvert\,\textless\,20^{\circ}$) prevents us from judging whether they contribute significantly to density cutoffs. 
However, the HAC and VOD partially contribute to the tilt of the halo, and the removal of them alleviates the deviation of the \textit{y}-axis, that is, $\Delta\gamma$ and $\Delta\eta\sim 0.08\,{\rm rad}$ or $4.58^{\circ}$.
Because their number density is not symmetrical with respect to the original \textit{y}-axis, the axis will tend to be biased in the direction of higher density.
We note that the orientation of the radially anisotropic halo is almost consistent with that of the entire galactic halo, especially $\beta$, and it has significant density cutoffs at 18 and 33\,kpc. Note that here the \textit{z}-axis is the major axis, but it is closer to the original \textit{x}-axis due to large $\beta$.
Therefore, we suggest that the radially anisotropic halo majorly contributes to two breaks and the tilt of the stellar halo.

\section{Conclusions}\label{Conclusions}

Based on the RRab stars sample, we apply a Bayesian approach to the Gaussian mixture model and find that the HAC and VOD have the radial anisotropic weights of $0.67^{+0.09}_{-0.07}$ and $0.57^{+0.07}_{-0.06}$, respectively, so they could be unmixed GSE debris.
Using 1,112 K giants with full 7D phase-space information, we study the substructures in the HAC and VOD.
Among the 568 and 544 K giants from HAC and VOD, we respectively identify 326 and 402 stars from the known substructures, including Sgr, GSE, Helmi Streams, Sequoia, Thamnos, Pontus, Wukong, and MWTD.
We identify stars from all the above substructures in the K giant sample of HAC, and it includes more retrograde substructures than VOD, but just a few Sgr stars exist in the HAC field.
As for VOD, except for the MWTD as well as the retrograde Pontus and Thamnos, stars from the other substructures have been discovered and tend to move prograde on polar orbits.
From HAC to VOD, the GSE becomes more prograde, and retrograde substructures, such as Pontus, Sequoia, and Thamnos, show similar behaviour, which means that they may have a common origin.
In the density fitting, the absence of the HAC and VOD structures alleviates the deviation of the \textit{y}-axis or the intermediate axis in the triaxial ellipsoid.
Finally, we find that the tilt and density cutoffs of the stellar halo are majorly contributed by the radially anisotropic halo.

\section*{Acknowledgements}
We thank the referee for the insightful comments and suggestions, which have improved the paper significantly. D.Y. thanks Chao Liu, Wenbo Wu and Giuliano Iorio for the helpful discussion.
This work was supported by the National Natural Sciences Foundation of China (NSFC Nos: 12090040, 12090044, 11973042, 11973052, and 11873053).  
It was also supported by the Fundamental Research Funds for the Central Universities and the National Key R\&D Program of China No. 2019YFA0405501. 
This work has made use of data from the European Space Agency (ESA) mission
{\it Gaia} (\url{https://www.cosmos.esa.int/gaia}), processed by the {\it Gaia}
Data Processing and Analysis Consortium (DPAC,
\url{https://www.cosmos.esa.int/web/gaia/dpac/consortium}). Funding for the DPAC
has been provided by national institutions, in particular the institutions
participating in the {\it Gaia} Multilateral Agreement.
Guoshoujing Telescope (the Large Sky Area Multi-Object Fiber Spectroscopic Telescope LAMOST) is a National Major Scientific Project built by the Chinese Academy of Sciences. Funding for the project has been provided by the National Development and Reform Commission. LAMOST is operated and managed by the National Astronomical Observatories, Chinese Academy of Sciences.

%%%%%%%%%%%%%%%%%%%%%%%%%%%%%%%%%%%%%%%%%%%%%%%%%%
\section*{Data Availability}

\textit{Gaia} DR3 and K giant data are publicly available.



\bibliographystyle{mnras}
\bibliography{yds3} % if your bibtex file is called example.bib

\begin{thebibliography}{}
\makeatletter
\relax
\def\mn@urlcharsother{\let\do\@makeother \do\$\do\&\do\#\do\^\do\_\do\%\do\~}
\def\mn@doi{\begingroup\mn@urlcharsother \@ifnextchar [ {\mn@doi@}
  {\mn@doi@[]}}
\def\mn@doi@[#1]#2{\def\@tempa{#1}\ifx\@tempa\@empty \href
  {http://dx.doi.org/#2} {doi:#2}\else \href {http://dx.doi.org/#2} {#1}\fi
  \endgroup}
\def\mn@eprint#1#2{\mn@eprint@#1:#2::\@nil}
\def\mn@eprint@arXiv#1{\href {http://arxiv.org/abs/#1} {{\tt arXiv:#1}}}
\def\mn@eprint@dblp#1{\href {http://dblp.uni-trier.de/rec/bibtex/#1.xml}
  {dblp:#1}}
\def\mn@eprint@#1:#2:#3:#4\@nil{\def\@tempa {#1}\def\@tempb {#2}\def\@tempc
  {#3}\ifx \@tempc \@empty \let \@tempc \@tempb \let \@tempb \@tempa \fi \ifx
  \@tempb \@empty \def\@tempb {arXiv}\fi \@ifundefined
  {mn@eprint@\@tempb}{\@tempb:\@tempc}{\expandafter \expandafter \csname
  mn@eprint@\@tempb\endcsname \expandafter{\@tempc}}}

\bibitem[\protect\citeauthoryear{{Abdurro'uf} et~al.,}{{Abdurro'uf}
  et~al.}{2022}]{Abdurro2022ApJS}
{Abdurro'uf} et~al., 2022, \mn@doi [\apjs] {10.3847/1538-4365/ac4414}, \href
  {https://ui.adsabs.harvard.edu/abs/2022ApJS..259...35A} {259, 35}

\bibitem[\protect\citeauthoryear{{Alam} et~al.,}{{Alam}
  et~al.}{2015}]{Alam2015ApJS}
{Alam} S.,  et~al., 2015, \mn@doi [\apjs] {10.1088/0067-0049/219/1/12}, \href
  {https://ui.adsabs.harvard.edu/abs/2015ApJS..219...12A} {219, 12}

\bibitem[\protect\citeauthoryear{{Amarante}, {Debattista}, {Beraldo e Silva},
  {Laporte}  \& {Deg}}{{Amarante} et~al.}{2022}]{Amarante2022ApJ}
{Amarante} J. A.~S.,  {Debattista} V.~P.,  {Beraldo e Silva} L.,  {Laporte} C.
  F.~P.,   {Deg} N.,  2022, \mn@doi [\apj] {10.3847/1538-4357/ac8b0d}, \href
  {https://ui.adsabs.harvard.edu/abs/2022ApJ...937...12A} {937, 12}

\bibitem[\protect\citeauthoryear{{Belokurov} et~al.,}{{Belokurov}
  et~al.}{2007}]{Belokurov2007ApJ}
{Belokurov} V.,  et~al., 2007, \mn@doi [\apjl] {10.1086/513144}, \href
  {https://ui.adsabs.harvard.edu/abs/2007ApJ...657L..89B} {657, L89}

\bibitem[\protect\citeauthoryear{{Belokurov} et~al.,}{{Belokurov}
  et~al.}{2014}]{Belokurov2014MNRAS}
{Belokurov} V.,  et~al., 2014, \mn@doi [\mnras] {10.1093/mnras/stt1862}, \href
  {https://ui.adsabs.harvard.edu/abs/2014MNRAS.437..116B} {437, 116}

\bibitem[\protect\citeauthoryear{{Belokurov}, {Erkal}, {Evans}, {Koposov}  \&
  {Deason}}{{Belokurov} et~al.}{2018}]{Belokurov2018MNRAS}
{Belokurov} V.,  {Erkal} D.,  {Evans} N.~W.,  {Koposov} S.~E.,   {Deason}
  A.~J.,  2018, \mn@doi [\mnras] {10.1093/mnras/sty982}, \href
  {https://ui.adsabs.harvard.edu/abs/2018MNRAS.478..611B} {478, 611}

\bibitem[\protect\citeauthoryear{{Belokurov}, {Sanders}, {Fattahi}, {Smith},
  {Deason}, {Evans}  \& {Grand}}{{Belokurov} et~al.}{2020}]{Belokurov2020MNRAS}
{Belokurov} V.,  {Sanders} J.~L.,  {Fattahi} A.,  {Smith} M.~C.,  {Deason}
  A.~J.,  {Evans} N.~W.,   {Grand} R. J.~J.,  2020, \mn@doi [\mnras]
  {10.1093/mnras/staa876}, \href
  {https://ui.adsabs.harvard.edu/abs/2020MNRAS.494.3880B} {494, 3880}

\bibitem[\protect\citeauthoryear{{Belokurov}, {Vasiliev}, {Deason}, {Koposov},
  {Fattahi}, {Dillamore}, {Davies}  \& {Grand}}{{Belokurov}
  et~al.}{2023}]{Belokurov2023MNRAS}
{Belokurov} V.,  {Vasiliev} E.,  {Deason} A.~J.,  {Koposov} S.~E.,  {Fattahi}
  A.,  {Dillamore} A.~M.,  {Davies} E.~Y.,   {Grand} R. J.~J.,  2023, \mn@doi
  [\mnras] {10.1093/mnras/stac3436}, \href
  {https://ui.adsabs.harvard.edu/abs/2023MNRAS.518.6200B} {518, 6200}

\bibitem[\protect\citeauthoryear{{Besla}, {Kallivayalil}, {Hernquist},
  {Robertson}, {Cox}, {van der Marel}  \& {Alcock}}{{Besla}
  et~al.}{2007}]{Besla2007ApJ}
{Besla} G.,  {Kallivayalil} N.,  {Hernquist} L.,  {Robertson} B.,  {Cox} T.~J.,
   {van der Marel} R.~P.,   {Alcock} C.,  2007, \mn@doi [\apj]
  {10.1086/521385}, \href
  {https://ui.adsabs.harvard.edu/abs/2007ApJ...668..949B} {668, 949}

\bibitem[\protect\citeauthoryear{{Binney}}{{Binney}}{2012}]{Binney2012MNRAS}
{Binney} J.,  2012, \mn@doi [\mnras] {10.1111/j.1365-2966.2012.21757.x}, \href
  {https://ui.adsabs.harvard.edu/abs/2012MNRAS.426.1324B} {426, 1324}

\bibitem[\protect\citeauthoryear{Binney \& Tremaine}{Binney \&
  Tremaine}{2008}]{binney2008st}
Binney J.,  Tremaine S.,  2008, Princeton Series in Astrophysics. 2nd ed.
  Princeton, NJ: Princeton University Press

\bibitem[\protect\citeauthoryear{{Bland-Hawthorn} \&
  {Gerhard}}{{Bland-Hawthorn} \& {Gerhard}}{2016}]{bland2016ARA&A}
{Bland-Hawthorn} J.,  {Gerhard} O.,  2016, \mn@doi [\araa]
  {10.1146/annurev-astro-081915-023441}, \href
  {https://ui.adsabs.harvard.edu/abs/2016ARA&A..54..529B} {54, 529}

\bibitem[\protect\citeauthoryear{{Blumenthal}, {Faber}, {Primack}  \&
  {Rees}}{{Blumenthal} et~al.}{1984}]{Blumenthal1984Natur}
{Blumenthal} G.~R.,  {Faber} S.~M.,  {Primack} J.~R.,   {Rees} M.~J.,  1984,
  \mn@doi [\nat] {10.1038/311517a0}, \href
  {https://ui.adsabs.harvard.edu/abs/1984Natur.311..517B} {311, 517}

\bibitem[\protect\citeauthoryear{{Bonaca} et~al.,}{{Bonaca}
  et~al.}{2012}]{Bonaca2012AJ}
{Bonaca} A.,  et~al., 2012, \mn@doi [\aj] {10.1088/0004-6256/143/5/105}, \href
  {https://ui.adsabs.harvard.edu/abs/2012AJ....143..105B} {143, 105}

\bibitem[\protect\citeauthoryear{{Boubert}, {Belokurov}, {Erkal}  \&
  {Iorio}}{{Boubert} et~al.}{2019}]{Boubert2019MNRAS}
{Boubert} D.,  {Belokurov} V.,  {Erkal} D.,   {Iorio} G.,  2019, \mn@doi
  [\mnras] {10.1093/mnras/sty3014}, \href
  {https://ui.adsabs.harvard.edu/abs/2019MNRAS.482.4562B} {482, 4562}

\bibitem[\protect\citeauthoryear{{Bovy}}{{Bovy}}{2015}]{Bovy2015ApJSgalpy}
{Bovy} J.,  2015, \mn@doi [\apjs] {10.1088/0067-0049/216/2/29}, \href
  {https://ui.adsabs.harvard.edu/abs/2015ApJS..216...29B} {216, 29}

\bibitem[\protect\citeauthoryear{{Bovy}, {Hogg}  \& {Roweis}}{{Bovy}
  et~al.}{2011}]{Bovy2011AnApS}
{Bovy} J.,  {Hogg} D.~W.,   {Roweis} S.~T.,  2011, \mn@doi [Annals of Applied
  Statistics] {10.1214/10-AOAS439}, \href
  {https://ui.adsabs.harvard.edu/abs/2011AnApS...5.1657B} {5, 1657}

\bibitem[\protect\citeauthoryear{{Buder} et~al.,}{{Buder}
  et~al.}{2021}]{Buder2021MNRAS}
{Buder} S.,  et~al., 2021, \mn@doi [\mnras] {10.1093/mnras/stab1242}, \href
  {https://ui.adsabs.harvard.edu/abs/2021MNRAS.506..150B} {506, 150}

\bibitem[\protect\citeauthoryear{{Bullock} \& {Johnston}}{{Bullock} \&
  {Johnston}}{2005}]{Bullock2005ApJ}
{Bullock} J.~S.,  {Johnston} K.~V.,  2005, \mn@doi [\apj] {10.1086/497422},
  \href {https://ui.adsabs.harvard.edu/abs/2005ApJ...635..931B} {635, 931}

\bibitem[\protect\citeauthoryear{{Cappellari} \& {Copin}}{{Cappellari} \&
  {Copin}}{2003}]{vorbin2003MNRAS}
{Cappellari} M.,  {Copin} Y.,  2003, \mn@doi [\mnras]
  {10.1046/j.1365-8711.2003.06541.x}, \href
  {https://ui.adsabs.harvard.edu/abs/2003MNRAS.342..345C} {342, 345}

\bibitem[\protect\citeauthoryear{{Carollo} et~al.,}{{Carollo}
  et~al.}{2019}]{Carollo2019ApJ}
{Carollo} D.,  et~al., 2019, \mn@doi [\apj] {10.3847/1538-4357/ab517c}, \href
  {https://ui.adsabs.harvard.edu/abs/2019ApJ...887...22C} {887, 22}

\bibitem[\protect\citeauthoryear{{Clementini} et~al.,}{{Clementini}
  et~al.}{2023}]{Clementini2023A&A}
{Clementini} G.,  et~al., 2023, \mn@doi [\aap] {10.1051/0004-6361/202243964},
  \href {https://ui.adsabs.harvard.edu/abs/2023A&A...674A..18C} {674, A18}

\bibitem[\protect\citeauthoryear{Cui et~al.,}{Cui et~al.}{2012}]{cui2012large}
Cui X.-Q.,  et~al., 2012, Research in Astronomy and Astrophysics, 12, 1197

\bibitem[\protect\citeauthoryear{{De Silva} et~al.,}{{De Silva}
  et~al.}{2015}]{GALAH2015MNRAS}
{De Silva} G.~M.,  et~al., 2015, \mn@doi [\mnras] {10.1093/mnras/stv327}, \href
  {https://ui.adsabs.harvard.edu/abs/2015MNRAS.449.2604D} {449, 2604}

\bibitem[\protect\citeauthoryear{{Deason}, {Belokurov}, {Koposov}  \&
  {Lancaster}}{{Deason} et~al.}{2018}]{Deason2018ApJ}
{Deason} A.~J.,  {Belokurov} V.,  {Koposov} S.~E.,   {Lancaster} L.,  2018,
  \mn@doi [\apjl] {10.3847/2041-8213/aad0ee}, \href
  {https://ui.adsabs.harvard.edu/abs/2018ApJ...862L...1D} {862, L1}

\bibitem[\protect\citeauthoryear{{Foreman-Mackey}, {Hogg}, {Lang}  \&
  {Goodman}}{{Foreman-Mackey} et~al.}{2013}]{Foreman2013PASP}
{Foreman-Mackey} D.,  {Hogg} D.~W.,  {Lang} D.,   {Goodman} J.,  2013, \mn@doi
  [\pasp] {10.1086/670067}, \href
  {https://ui.adsabs.harvard.edu/abs/2013PASP..125..306F} {125, 306}

\bibitem[\protect\citeauthoryear{{Gaia Collaboration} et~al.,}{{Gaia
  Collaboration} et~al.}{2016}]{Gaia2016A&A}
{Gaia Collaboration} et~al., 2016, \mn@doi [\aap]
  {10.1051/0004-6361/201629272}, \href
  {https://ui.adsabs.harvard.edu/abs/2016A&A...595A...1G} {595, A1}

\bibitem[\protect\citeauthoryear{{Gaia Collaboration} et~al.,}{{Gaia
  Collaboration} et~al.}{2023}]{GaiaDR32023A&A}
{Gaia Collaboration} et~al., 2023, \mn@doi [\aap]
  {10.1051/0004-6361/202243940}, \href
  {https://ui.adsabs.harvard.edu/abs/2023A&A...674A...1G} {674, A1}

\bibitem[\protect\citeauthoryear{{Gallart}, {Bernard}, {Brook}, {Ruiz-Lara},
  {Cassisi}, {Hill}  \& {Monelli}}{{Gallart} et~al.}{2019}]{Gallart2019NatAs}
{Gallart} C.,  {Bernard} E.~J.,  {Brook} C.~B.,  {Ruiz-Lara} T.,  {Cassisi} S.,
   {Hill} V.,   {Monelli} M.,  2019, \mn@doi [Nature Astronomy]
  {10.1038/s41550-019-0829-5}, \href
  {https://ui.adsabs.harvard.edu/abs/2019NatAs...3..932G} {3, 932}

\bibitem[\protect\citeauthoryear{Goodman \& Weare}{Goodman \&
  Weare}{2010}]{goodman2010ensemble}
Goodman J.,  Weare J.,  2010, Communications in applied mathematics and
  computational science, 5, 65

\bibitem[\protect\citeauthoryear{{Han} et~al.,}{{Han}
  et~al.}{2022a}]{Han2022AJ}
{Han} J.~J.,  et~al., 2022a, \mn@doi [\aj] {10.3847/1538-3881/ac97e9}, \href
  {https://ui.adsabs.harvard.edu/abs/2022AJ....164..249H} {164, 249}

\bibitem[\protect\citeauthoryear{{Han} et~al.,}{{Han}
  et~al.}{2022b}]{Han2022ApJ}
{Han} J.~J.,  et~al., 2022b, \mn@doi [\apj] {10.3847/1538-4357/ac795f}, \href
  {https://ui.adsabs.harvard.edu/abs/2022ApJ...934...14H} {934, 14}

\bibitem[\protect\citeauthoryear{{Harris}}{{Harris}}{1996}]{Harris1996J}
{Harris} W.~E.,  1996, \mn@doi [\aj] {10.1086/118116}, \href
  {https://ui.adsabs.harvard.edu/abs/1996AJ....112.1487H} {112, 1487}

\bibitem[\protect\citeauthoryear{{Haywood}, {Di Matteo}, {Lehnert}, {Snaith},
  {Khoperskov}  \& {G{\'o}mez}}{{Haywood} et~al.}{2018}]{Haywood2018ApJ}
{Haywood} M.,  {Di Matteo} P.,  {Lehnert} M.~D.,  {Snaith} O.,  {Khoperskov}
  S.,   {G{\'o}mez} A.,  2018, \mn@doi [\apj] {10.3847/1538-4357/aad235}, \href
  {https://ui.adsabs.harvard.edu/abs/2018ApJ...863..113H} {863, 113}

\bibitem[\protect\citeauthoryear{{Helmi}, {Babusiaux}, {Koppelman}, {Massari},
  {Veljanoski}  \& {Brown}}{{Helmi} et~al.}{2018}]{Helmi2018Natur}
{Helmi} A.,  {Babusiaux} C.,  {Koppelman} H.~H.,  {Massari} D.,  {Veljanoski}
  J.,   {Brown} A. G.~A.,  2018, \mn@doi [\nat] {10.1038/s41586-018-0625-x},
  \href {https://ui.adsabs.harvard.edu/abs/2018Natur.563...85H} {563, 85}

\bibitem[\protect\citeauthoryear{{Hernitschek} et~al.,}{{Hernitschek}
  et~al.}{2018}]{Hernitschek2018ApJ}
{Hernitschek} N.,  et~al., 2018, \mn@doi [\apj] {10.3847/1538-4357/aabfbb},
  \href {https://ui.adsabs.harvard.edu/abs/2018ApJ...859...31H} {859, 31}

\bibitem[\protect\citeauthoryear{{Horta} et~al.,}{{Horta}
  et~al.}{2023}]{Horta2023MNRAS}
{Horta} D.,  et~al., 2023, \mn@doi [\mnras] {10.1093/mnras/stac3179}, \href
  {https://ui.adsabs.harvard.edu/abs/2023MNRAS.520.5671H} {520, 5671}

\bibitem[\protect\citeauthoryear{{Huang} et~al.,}{{Huang}
  et~al.}{2021}]{Huang2021ApJ}
{Huang} Y.,  et~al., 2021, \mn@doi [\apj] {10.3847/1538-4357/abca37}, \href
  {https://ui.adsabs.harvard.edu/abs/2021ApJ...907...68H} {907, 68}

\bibitem[\protect\citeauthoryear{{Iorio} \& {Belokurov}}{{Iorio} \&
  {Belokurov}}{2019}]{Iorio2019MNRAS}
{Iorio} G.,  {Belokurov} V.,  2019, \mn@doi [\mnras] {10.1093/mnras/sty2806},
  \href {https://ui.adsabs.harvard.edu/abs/2019MNRAS.482.3868I} {482, 3868}

\bibitem[\protect\citeauthoryear{{Iorio} \& {Belokurov}}{{Iorio} \&
  {Belokurov}}{2021}]{iorio2021MNRAS}
{Iorio} G.,  {Belokurov} V.,  2021, \mn@doi [\mnras] {10.1093/mnras/stab005},
  \href {https://ui.adsabs.harvard.edu/abs/2021MNRAS.502.5686I} {502, 5686}

\bibitem[\protect\citeauthoryear{{Iorio}, {Belokurov}, {Erkal}, {Koposov},
  {Nipoti}  \& {Fraternali}}{{Iorio} et~al.}{2018}]{Iorio2018MNRAS}
{Iorio} G.,  {Belokurov} V.,  {Erkal} D.,  {Koposov} S.~E.,  {Nipoti} C.,
  {Fraternali} F.,  2018, \mn@doi [\mnras] {10.1093/mnras/stx2819}, \href
  {https://ui.adsabs.harvard.edu/abs/2018MNRAS.474.2142I} {474, 2142}

\bibitem[\protect\citeauthoryear{{Ivezi{\'c}}, {Connolly}, {Vanderplas}  \&
  {Gray}}{{Ivezi{\'c}} et~al.}{2014}]{astroMLText}
{Ivezi{\'c}} {\v Z}.,  {Connolly} A.,  {Vanderplas} J.,   {Gray} A.,  2014,
  Statistics, Data Mining and Machine Learning in Astronomy.
Princeton University Press

\bibitem[\protect\citeauthoryear{{Johnson} \& {Soderblom}}{{Johnson} \&
  {Soderblom}}{1987}]{johnson1987AJ}
{Johnson} D. R.~H.,  {Soderblom} D.~R.,  1987, \mn@doi [\aj] {10.1086/114370},
  \href {https://ui.adsabs.harvard.edu/abs/1987AJ.....93..864J} {93, 864}

\bibitem[\protect\citeauthoryear{{Juri{\'c}} et~al.,}{{Juri{\'c}}
  et~al.}{2008}]{Juri2008ApJ}
{Juri{\'c}} M.,  et~al., 2008, \mn@doi [\apj] {10.1086/523619}, \href
  {https://ui.adsabs.harvard.edu/abs/2008ApJ...673..864J} {673, 864}

\bibitem[\protect\citeauthoryear{{Koppelman}, {Helmi}, {Massari},
  {Price-Whelan}  \& {Starkenburg}}{{Koppelman}
  et~al.}{2019}]{Koppelman2019A&A}
{Koppelman} H.~H.,  {Helmi} A.,  {Massari} D.,  {Price-Whelan} A.~M.,
  {Starkenburg} T.~K.,  2019, \mn@doi [\aap] {10.1051/0004-6361/201936738},
  \href {https://ui.adsabs.harvard.edu/abs/2019A&A...631L...9K} {631, L9}

\bibitem[\protect\citeauthoryear{{Kruijssen}, {Pfeffer}, {Reina-Campos},
  {Crain}  \& {Bastian}}{{Kruijssen} et~al.}{2019}]{Kruijssen2019MNRAS}
{Kruijssen} J.~M.~D.,  {Pfeffer} J.~L.,  {Reina-Campos} M.,  {Crain} R.~A.,
  {Bastian} N.,  2019, \mn@doi [\mnras] {10.1093/mnras/sty1609}, \href
  {https://ui.adsabs.harvard.edu/abs/2019MNRAS.486.3180K} {486, 3180}

\bibitem[\protect\citeauthoryear{{Lancaster}, {Koposov}, {Belokurov}, {Evans}
  \& {Deason}}{{Lancaster} et~al.}{2019}]{lancaster2019MNRAS}
{Lancaster} L.,  {Koposov} S.~E.,  {Belokurov} V.,  {Evans} N.~W.,   {Deason}
  A.~J.,  2019, \mn@doi [\mnras] {10.1093/mnras/stz853}, \href
  {https://ui.adsabs.harvard.edu/abs/2019MNRAS.486..378L} {486, 378}

\bibitem[\protect\citeauthoryear{{Law} \& {Majewski}}{{Law} \&
  {Majewski}}{2010}]{Law2010ApJ}
{Law} D.~R.,  {Majewski} S.~R.,  2010, \mn@doi [\apj]
  {10.1088/0004-637X/714/1/229}, \href
  {https://ui.adsabs.harvard.edu/abs/2010ApJ...714..229L} {714, 229}

\bibitem[\protect\citeauthoryear{{Li}, {Huang}, {Liu}, {Beers}  \&
  {Zhang}}{{Li} et~al.}{2023}]{Lixinyi2023ApJ}
{Li} X.-Y.,  {Huang} Y.,  {Liu} G.-C.,  {Beers} T.~C.,   {Zhang} H.-W.,  2023,
  \mn@doi [\apj] {10.3847/1538-4357/acadd5}, \href
  {https://ui.adsabs.harvard.edu/abs/2023ApJ...944...88L} {944, 88}

\bibitem[\protect\citeauthoryear{{Majewski} et~al.,}{{Majewski}
  et~al.}{2017}]{apogee2017AJ}
{Majewski} S.~R.,  et~al., 2017, \mn@doi [\aj] {10.3847/1538-3881/aa784d},
  \href {https://ui.adsabs.harvard.edu/abs/2017AJ....154...94M} {154, 94}

\bibitem[\protect\citeauthoryear{{Malhan} et~al.,}{{Malhan}
  et~al.}{2022}]{Malhan2022ApJ}
{Malhan} K.,  et~al., 2022, \mn@doi [\apj] {10.3847/1538-4357/ac4d2a}, \href
  {https://ui.adsabs.harvard.edu/abs/2022ApJ...926..107M} {926, 107}

\bibitem[\protect\citeauthoryear{{Martell} et~al.,}{{Martell}
  et~al.}{2017}]{Martell2017MNRAS}
{Martell} S.~L.,  et~al., 2017, \mn@doi [\mnras] {10.1093/mnras/stw2835}, \href
  {https://ui.adsabs.harvard.edu/abs/2017MNRAS.465.3203M} {465, 3203}

\bibitem[\protect\citeauthoryear{{Mateu}}{{Mateu}}{2023}]{Mateu2023MNRAS}
{Mateu} C.,  2023, \mn@doi [\mnras] {10.1093/mnras/stad321}, \href
  {https://ui.adsabs.harvard.edu/abs/2023MNRAS.520.5225M} {520, 5225}

\bibitem[\protect\citeauthoryear{{Mateu}}{{Mateu}}{2024}]{Mateu2024RNAAS}
{Mateu} C.,  2024, \mn@doi [Research Notes of the American Astronomical
  Society] {10.3847/2515-5172/ad3540}, \href
  {https://ui.adsabs.harvard.edu/abs/2024RNAAS...8...85M} {8, 85}

\bibitem[\protect\citeauthoryear{{McMillan}}{{McMillan}}{2017}]{McMillan2017MNRAS}
{McMillan} P.~J.,  2017, \mn@doi [\mnras] {10.1093/mnras/stw2759}, \href
  {https://ui.adsabs.harvard.edu/abs/2017MNRAS.465...76M} {465, 76}

\bibitem[\protect\citeauthoryear{{Myeong}, {Evans}, {Belokurov}, {Sanders}  \&
  {Koposov}}{{Myeong} et~al.}{2018a}]{Myeong2018ApJ1}
{Myeong} G.~C.,  {Evans} N.~W.,  {Belokurov} V.,  {Sanders} J.~L.,   {Koposov}
  S.~E.,  2018a, \mn@doi [\apjl] {10.3847/2041-8213/aab613}, \href
  {https://ui.adsabs.harvard.edu/abs/2018ApJ...856L..26M} {856, L26}

\bibitem[\protect\citeauthoryear{{Myeong}, {Evans}, {Belokurov}, {Sanders}  \&
  {Koposov}}{{Myeong} et~al.}{2018b}]{Myeong2018ApJ2}
{Myeong} G.~C.,  {Evans} N.~W.,  {Belokurov} V.,  {Sanders} J.~L.,   {Koposov}
  S.~E.,  2018b, \mn@doi [\apjl] {10.3847/2041-8213/aad7f7}, \href
  {https://ui.adsabs.harvard.edu/abs/2018ApJ...863L..28M} {863, L28}

\bibitem[\protect\citeauthoryear{{Myeong}, {Vasiliev}, {Iorio}, {Evans}  \&
  {Belokurov}}{{Myeong} et~al.}{2019}]{Myeong2019MNRAS}
{Myeong} G.~C.,  {Vasiliev} E.,  {Iorio} G.,  {Evans} N.~W.,   {Belokurov} V.,
  2019, \mn@doi [\mnras] {10.1093/mnras/stz1770}, \href
  {https://ui.adsabs.harvard.edu/abs/2019MNRAS.488.1235M} {488, 1235}

\bibitem[\protect\citeauthoryear{{Naidu}, {Conroy}, {Bonaca}, {Johnson},
  {Ting}, {Caldwell}, {Zaritsky}  \& {Cargile}}{{Naidu}
  et~al.}{2020}]{Naidu2020ApJ}
{Naidu} R.~P.,  {Conroy} C.,  {Bonaca} A.,  {Johnson} B.~D.,  {Ting} Y.-S.,
  {Caldwell} N.,  {Zaritsky} D.,   {Cargile} P.~A.,  2020, \mn@doi [\apj]
  {10.3847/1538-4357/abaef4}, \href
  {https://ui.adsabs.harvard.edu/abs/2020ApJ...901...48N} {901, 48}

\bibitem[\protect\citeauthoryear{{Naidu} et~al.,}{{Naidu}
  et~al.}{2021}]{naidu2021ApJ}
{Naidu} R.~P.,  et~al., 2021, \mn@doi [\apj] {10.3847/1538-4357/ac2d2d}, \href
  {https://ui.adsabs.harvard.edu/abs/2021ApJ...923...92N} {923, 92}

\bibitem[\protect\citeauthoryear{{Newberg} et~al.,}{{Newberg}
  et~al.}{2002}]{Newberg2002ApJ}
{Newberg} H.~J.,  et~al., 2002, \mn@doi [\apj] {10.1086/338983}, \href
  {https://ui.adsabs.harvard.edu/abs/2002ApJ...569..245N} {569, 245}

\bibitem[\protect\citeauthoryear{{Perottoni}, {Limberg}, {Amarante}, {Rossi},
  {Queiroz}, {Santucci}, {P{\'e}rez-Villegas}  \& {Chiappini}}{{Perottoni}
  et~al.}{2022}]{Perottoni2022ApJ}
{Perottoni} H.~D.,  {Limberg} G.,  {Amarante} J. A.~S.,  {Rossi} S.,  {Queiroz}
  A. B.~A.,  {Santucci} R.~M.,  {P{\'e}rez-Villegas} A.,   {Chiappini} C.,
  2022, \mn@doi [\apjl] {10.3847/2041-8213/ac88d6}, \href
  {https://ui.adsabs.harvard.edu/abs/2022ApJ...936L...2P} {936, L2}

\bibitem[\protect\citeauthoryear{{Schlegel}, {Finkbeiner}  \&
  {Davis}}{{Schlegel} et~al.}{1998}]{SFD1998ApJ}
{Schlegel} D.~J.,  {Finkbeiner} D.~P.,   {Davis} M.,  1998, \mn@doi [\apj]
  {10.1086/305772}, \href
  {https://ui.adsabs.harvard.edu/abs/1998ApJ...500..525S} {500, 525}

\bibitem[\protect\citeauthoryear{{Searle} \& {Zinn}}{{Searle} \&
  {Zinn}}{1978}]{Searle1978ApJ}
{Searle} L.,  {Zinn} R.,  1978, \mn@doi [\apj] {10.1086/156499}, \href
  {https://ui.adsabs.harvard.edu/abs/1978ApJ...225..357S} {225, 357}

\bibitem[\protect\citeauthoryear{{Simion}, {Belokurov}  \& {Koposov}}{{Simion}
  et~al.}{2019}]{Simion2019MNRAS}
{Simion} I.~T.,  {Belokurov} V.,   {Koposov} S.~E.,  2019, \mn@doi [\mnras]
  {10.1093/mnras/sty2744}, \href
  {https://ui.adsabs.harvard.edu/abs/2019MNRAS.482..921S} {482, 921}

\bibitem[\protect\citeauthoryear{{Tian} et~al.,}{{Tian}
  et~al.}{2015}]{tian2015ApJ}
{Tian} H.-J.,  et~al., 2015, \mn@doi [\apj] {10.1088/0004-637X/809/2/145},
  \href {https://ui.adsabs.harvard.edu/abs/2015ApJ...809..145T} {809, 145}

\bibitem[\protect\citeauthoryear{VanderPlas, Connolly, Ivezi{\'c}  \&
  Gray}{VanderPlas et~al.}{2012}]{vanderplas2012introduction}
VanderPlas J.,  Connolly A.~J.,  Ivezi{\'c} {\v{Z}}.,   Gray A.,  2012, in 2012
  conference on intelligent data understanding. pp 47--54

\bibitem[\protect\citeauthoryear{{Vanderplas}, {Connolly}, {Ivezi{\'c}}  \&
  {Gray}}{{Vanderplas} et~al.}{2012}]{astroML}
{Vanderplas} J.,  {Connolly} A.,  {Ivezi{\'c}} {\v Z}.,   {Gray} A.,  2012, in
  Conference on Intelligent Data Understanding (CIDU). pp 47 --54,
  \mn@doi{10.1109/CIDU.2012.6382200}

\bibitem[\protect\citeauthoryear{{Vivas} et~al.,}{{Vivas}
  et~al.}{2001}]{Vivas2001ApJ}
{Vivas} A.~K.,  et~al., 2001, \mn@doi [\apjl] {10.1086/320915}, \href
  {https://ui.adsabs.harvard.edu/abs/2001ApJ...554L..33V} {554, L33}

\bibitem[\protect\citeauthoryear{{Wang}, {Zhang}, {Xue}, {Huang}, {Liu},
  {Zhang}  \& {Yang}}{{Wang} et~al.}{2022}]{Wang2022MNRAS}
{Wang} F.,  {Zhang} H.~W.,  {Xue} X.~X.,  {Huang} Y.,  {Liu} G.~C.,  {Zhang}
  L.,   {Yang} C.~Q.,  2022, \mn@doi [\mnras] {10.1093/mnras/stac874}, \href
  {https://ui.adsabs.harvard.edu/abs/2022MNRAS.513.1958W} {513, 1958}

\bibitem[\protect\citeauthoryear{{Wegg}, {Gerhard}  \& {Bieth}}{{Wegg}
  et~al.}{2019}]{Wegg2019MNRAS}
{Wegg} C.,  {Gerhard} O.,   {Bieth} M.,  2019, \mn@doi [\mnras]
  {10.1093/mnras/stz572}, \href
  {https://ui.adsabs.harvard.edu/abs/2019MNRAS.485.3296W} {485, 3296}

\bibitem[\protect\citeauthoryear{{Wu}, {Zhao}, {Xue}, {Bird}  \& {Yang}}{{Wu}
  et~al.}{2022}]{Wu2022ApJ}
{Wu} W.,  {Zhao} G.,  {Xue} X.-X.,  {Bird} S.~A.,   {Yang} C.,  2022, \mn@doi
  [\apj] {10.3847/1538-4357/ac31ac}, \href
  {https://ui.adsabs.harvard.edu/abs/2022ApJ...924...23W} {924, 23}

\bibitem[\protect\citeauthoryear{{Xue} et~al.,}{{Xue}
  et~al.}{2014}]{Xue2014ApJ}
{Xue} X.-X.,  et~al., 2014, \mn@doi [\apj] {10.1088/0004-637X/784/2/170}, \href
  {https://ui.adsabs.harvard.edu/abs/2014ApJ...784..170X} {784, 170}

\bibitem[\protect\citeauthoryear{{Xue}, {Rix}, {Ma}, {Morrison}, {Bovy},
  {Sesar}  \& {Janesh}}{{Xue} et~al.}{2015}]{Xue2015ApJ}
{Xue} X.-X.,  {Rix} H.-W.,  {Ma} Z.,  {Morrison} H.,  {Bovy} J.,  {Sesar} B.,
  {Janesh} W.,  2015, \mn@doi [\apj] {10.1088/0004-637X/809/2/144}, \href
  {https://ui.adsabs.harvard.edu/abs/2015ApJ...809..144X} {809, 144}

\bibitem[\protect\citeauthoryear{{Yan}, {Shi}, {Chen}, {Zhao}  \& {Zhao}}{{Yan}
  et~al.}{2023}]{Yan2023A&A}
{Yan} H.~H.,  {Shi} W.~B.,  {Chen} Y.~Q.,  {Zhao} J.~K.,   {Zhao} G.,  2023,
  \mn@doi [\aap] {10.1051/0004-6361/202346249}, \href
  {https://ui.adsabs.harvard.edu/abs/2023A&A...674A..78Y} {674, A78}

\bibitem[\protect\citeauthoryear{{Yanny} et~al.,}{{Yanny}
  et~al.}{2009}]{segue2009AJ}
{Yanny} B.,  et~al., 2009, \mn@doi [\aj] {10.1088/0004-6256/137/5/4377}, \href
  {https://ui.adsabs.harvard.edu/abs/2009AJ....137.4377Y} {137, 4377}

\bibitem[\protect\citeauthoryear{{Ye}, {Du}, {Shi}  \& {Ma}}{{Ye}
  et~al.}{2023}]{Ye2023MNRAS}
{Ye} D.,  {Du} C.,  {Shi} J.,   {Ma} J.,  2023, \mn@doi [\mnras]
  {10.1093/mnras/stad2320}, \href
  {https://ui.adsabs.harvard.edu/abs/2023MNRAS.525.2472Y} {525, 2472}

\bibitem[\protect\citeauthoryear{{Ye}, {Du}, {Shi}  \& {Ma}}{{Ye}
  et~al.}{2024}]{Ye2024}
{Ye} D.,  {Du} C.,  {Shi} J.,   {Ma} J.,  2024, \mn@doi [\mnras]
  {10.1093/mnras/stad3860}, \href
  {https://ui.adsabs.harvard.edu/abs/2024MNRAS.527.9892Y} {527, 9892}

\bibitem[\protect\citeauthoryear{{Yuan}, {Chang}, {Beers}  \& {Huang}}{{Yuan}
  et~al.}{2020}]{Yuan2020ApJ}
{Yuan} Z.,  {Chang} J.,  {Beers} T.~C.,   {Huang} Y.,  2020, \mn@doi [\apjl]
  {10.3847/2041-8213/aba49f}, \href
  {https://ui.adsabs.harvard.edu/abs/2020ApJ...898L..37Y} {898, L37}

\bibitem[\protect\citeauthoryear{{Zhang}, {Xue}, {Yang}, {Wang}, {Rix}, {Zhao}
  \& {Liu}}{{Zhang} et~al.}{2023}]{zhanglan2023}
{Zhang} L.,  {Xue} X.-X.,  {Yang} C.,  {Wang} F.,  {Rix} H.-W.,  {Zhao} G.,
  {Liu} C.,  2023, \mn@doi [\aj] {10.3847/1538-3881/acc9bb}, \href
  {https://ui.adsabs.harvard.edu/abs/2023AJ....165..224Z} {165, 224}

\bibitem[\protect\citeauthoryear{{Zhao}, {Chen}, {Shi}, {Liang}, {Hou}, {Chen},
  {Zhang}  \& {Li}}{{Zhao} et~al.}{2006}]{Zhao2006ChJAA}
{Zhao} G.,  {Chen} Y.-Q.,  {Shi} J.-R.,  {Liang} Y.-C.,  {Hou} J.-L.,  {Chen}
  L.,  {Zhang} H.-W.,   {Li} A.-G.,  2006, \mn@doi [\cjaa]
  {10.1088/1009-9271/6/3/01}, \href
  {https://ui.adsabs.harvard.edu/abs/2006ChJAA...6..265Z} {6, 265}

\bibitem[\protect\citeauthoryear{Zhao, Zhao, Chu, Jing  \& Deng}{Zhao
  et~al.}{2012}]{zhao2012lamost}
Zhao G.,  Zhao Y.-H.,  Chu Y.-Q.,  Jing Y.-P.,   Deng L.-C.,  2012, Research in
  Astronomy and Astrophysics, 12, 723

\bibitem[\protect\citeauthoryear{{van der Marel} \& {Sahlmann}}{{van der Marel}
  \& {Sahlmann}}{2016}]{van2016ApJ}
{van der Marel} R.~P.,  {Sahlmann} J.,  2016, \mn@doi [\apjl]
  {10.3847/2041-8205/832/2/L23}, \href
  {https://ui.adsabs.harvard.edu/abs/2016ApJ...832L..23V} {832, L23}

\makeatother
\end{thebibliography}

%\begin{thebibliography}{99}
%\end{thebibliography}

%%%%%%%%%%%%%%%%%%%%%%%%%%%%%%%%%%%%%%%%%%%%%%%%%%

%%%%%%%%%%%%%%%%% APPENDICES %%%%%%%%%%%%%%%%%%%%%

\appendix

%%%%%%%%%%%%%%%%%%%%%%%%%%%%%%%%%%%%%%%%%%%%%%%%%%

% Don't change these lines
\bsp	% typesetting comment
\label{lastpage}
\end{document}